\documentclass[conference]{IEEEtran}
\usepackage{fancyhdr}
\usepackage[utf8]{inputenc}
\IEEEoverridecommandlockouts
\usepackage{cite}
\usepackage{amsmath,amssymb,amsfonts}
\usepackage{graphicx}
\usepackage{textcomp}
\usepackage{xcolor}
\usepackage{algorithm}
\usepackage{algpseudocode}
\usepackage{booktabs}
\usepackage{multirow}
\usepackage{adjustbox}
\usepackage[caption=false,font=footnotesize]{subfig}
\usepackage{hyperref}
\usepackage{newunicodechar}
\usepackage{stfloats}
\newunicodechar{，}{,}
\newcommand{\sys}{\textsc{NCCLZ }}
\def\BibTeX{{\rm B\kern-.05em{\sc i\kern-.025em b}\kern-.08em
    T\kern-.1667em\lower.7ex\hbox{E}\kern-.125emX}}
\begin{document}

\title{NCCLZ: Compression-Enabled GPU Collectives with Decoupled Quantization and Entropy Coding
}

\author{
\IEEEauthorblockN{Jiamin Wang}
\IEEEauthorblockA{\textit{Department of Computer Science} \\
\textit{Stevens Institute of Technology}\\
Hoboken, USA \\
jwang259@stevens.edu}
\and
\IEEEauthorblockN{Zhijing Ye}
\IEEEauthorblockA{\textit{Department of Computer Science} \\
\textit{Stevens Institute of Technology}\\
Hoboken, USA \\
zye25@stevens.edu}
\and
\IEEEauthorblockN{Xiaodong Yu}
\IEEEauthorblockA{\textit{Department of Computer Science} \\
\textit{Stevens Institute of Technology}\\
Hoboken, USA \\
xyu38@stevens.edu}
}

\maketitle
\thispagestyle{fancy} 
\lhead{} 
\rhead{} 
\chead{} 
\lfoot{\footnotesize{ SC26, November 15-20, 2026, Chicago, Illinois, USA 
\newline 979-8-3195-4789-7/26/\$31.00 \copyright 2026 IEEE}} 
\rfoot{} 
\cfoot{} 
\renewcommand{\headrulewidth}{0pt} \renewcommand{\footrulewidth}{0pt} 

\begin{abstract}
Collective communication is a major bottleneck for multi-node GPU workloads in scientific computing and distributed deep learning, especially when inter-node bandwidth is limited. Although NCCL provides optimized GPU-centric collectives, large messages can still dominate end-to-end performance. Existing compression-enabled collective libraries either operate in MPI-based stacks rather than NCCL, omit entropy coding, or couple complete compressors to communication primitives; these choices can limit flexibility, compression ratio, and compression--communication overlap. This paper presents NCCLZ, a compression-enabled GPU collective framework that decouples quantization from entropy coding and integrates the two stages at different layers of the stack. NCCLZ places quantization at the interface, embeds entropy coding in NCCL primitives, uses a lightweight device-side selector to choose among coding strategies, and overlaps compression with communication to reduce exposed overhead. Experiments on scientific datasets, training gradients, and synthetic workloads show speedups of up to $9.65\times$ over NCCL and improvements of up to $3.34\times$ over prior compression-assisted collective libraries.
\end{abstract}


\section{Introduction}

GPU accelerators play a central role in modern HPC clusters, where scientific applications and distributed deep learning increasingly use multi-GPU execution across nodes. Performance in these systems is often limited by inter-GPU communication, particularly when inter-node bandwidth is constrained. For example, data-parallel training synchronizes gradients through collectives such as \texttt{AllReduce}~\cite{Sergeev2018Horovod} in every iteration. As accelerator throughput and model sizes increase, this synchronization can become a dominant bottleneck at scale~\cite{Zhang2020NetBottleneck,Poseidon2017ATC}. On NVIDIA platforms, NCCL provides GPU-native collective abstractions and topology-aware optimizations for interconnects such as NVLink. NCCL is therefore a widely used backend for multi-GPU systems, but its large-message inter-node performance remains limited by network bandwidth.

%Modern HPC clusters are increasingly dominated by NVIDIA GPU-based systems, where both large-scale scientific applications and distributed deep learning workloads rely on many-GPU computation across multiple compute nodes. In such environments, overall performance is often determined by communication efficiency among GPUs, particularly in inter-node and inter-rack settings where network bandwidth is limited. For example, in data-parallel deep learning training, each iteration produces gradients that must be synchronized across workers via collectives such as \texttt{AllReduce}~\cite{Sergeev2018Horovod}. As GPUs process more batches per unit time, synchronization becomes more frequent and message sizes continue to grow, pushing larger volumes of data into the network and making communication a dominant cost at scale~\cite{Zhang2020NetBottleneck,Poseidon2017ATC}. To support efficient GPU communication, NVIDIA provides NCCL, a vendor-optimized collective communication library that exposes GPU-native abstractions and incorporates topology-aware optimizations for NVLinks. Compared to general GPU-aware MPI, NCCL more effectively exploits NVIDIA GPU features and capabilities, and has therefore become the de facto communication backend for modern multi-GPU systems. However, while NCCL achieves near-device-level throughput in intra-node environments with NVLinks, its performance with large message sizes can still be constrained in inter-node scenarios where network bandwidth becomes the limiting factor.

\begin{figure*}
    \centering
\includegraphics[width=\linewidth
]{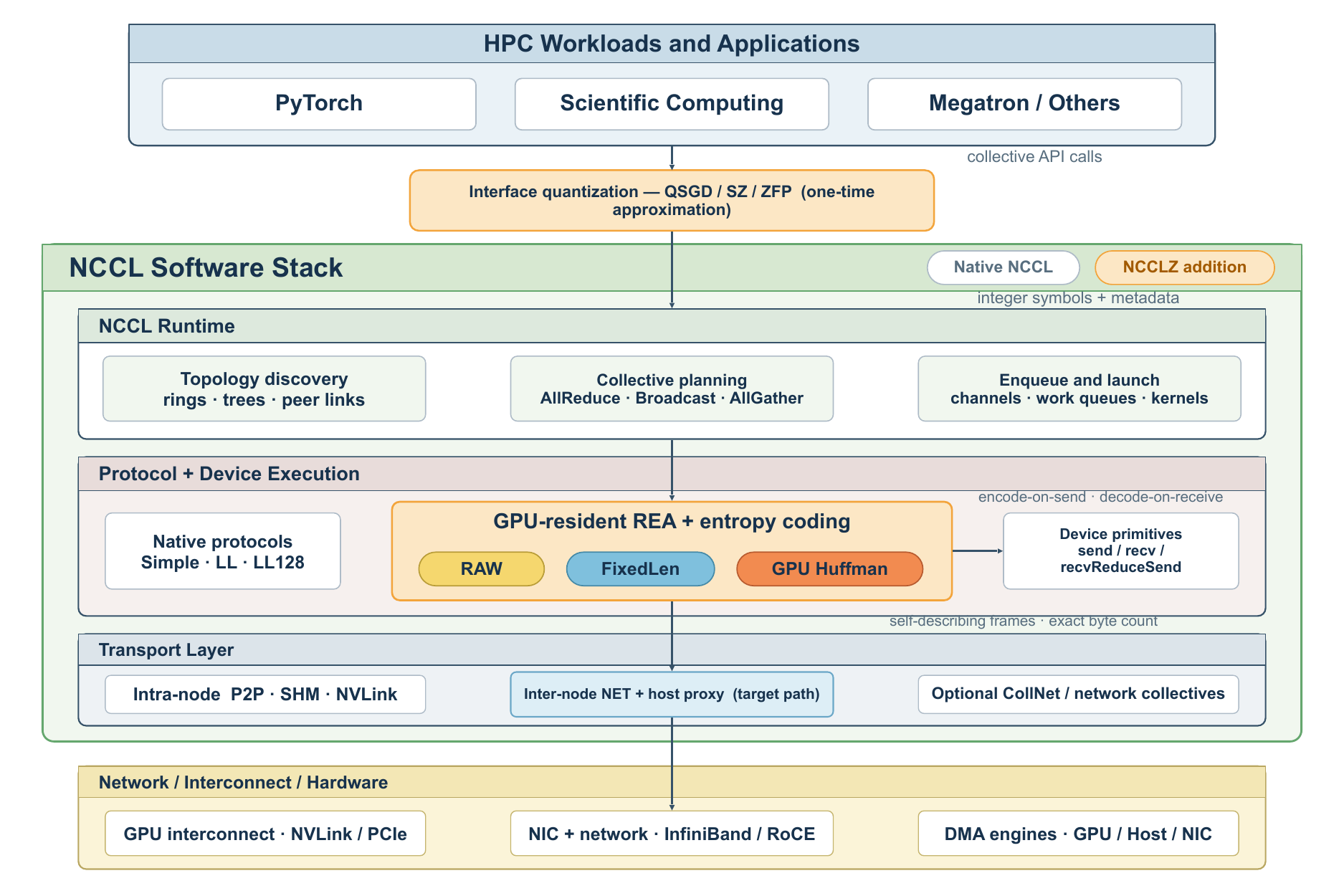}
\caption{NCCLZ system architecture. Quantization is applied at the interface, GPU-resident entropy coding is integrated into NCCL primitives, and NCCL's transport layer moves the resulting frames across the network.}
\vspace{-8pt}
    \label{fig:arch}
\end{figure*}

Lossy compression is widely used in scientific and AI workloads to reduce communication volume. Scientific compressors commonly combine lossy quantization with lossless entropy coding to maximize compression ratios~\cite{di2016sz,tian2020cusz}. By contrast, communication compression for modern AI workloads, especially LLMs, often uses quantization alone because entropy coding can incur substantial computational overhead~\cite{Chen2025GreedyLore,Feng2024DLRMCompression,He2025TAHQuant}. Entropy coding can still reduce quantized payloads, but it improves end-to-end performance only when its cost is sufficiently overlapped with communication.

Several systems integrate lossy compression into collective communication libraries (CCLs) to reduce network traffic while hiding compression overhead. MVAPICH2~\cite{Zhou2021MVAPICHCompression} supports on-the-fly compression in MPI collectives. gZCCL~\cite{Huang2024gZCCL} and ghZCCL~\cite{Huang2025ghZCCL} combine GPU-aware MPI collectives with GPU-based lossy and homomorphic compressors, respectively, and overlap compression with communication. More recently, COCCL~\cite{Liu2026COCCL} has integrated lightweight quantization into NCCL APIs for compression-accelerated collectives on NVIDIA GPUs.

These systems establish that compression, runtime codec selection, and GPU acceleration can benefit collective communication, including MPI-based collectives. However, their integration boundaries differ from the separation explored in this work. COCCL applies quantization at the NCCL API layer but does not include a separate entropy-coding stage. The cited MPI-based systems integrate and select complete compressors within an MPI-oriented communication path. Consequently, enabling entropy coding independently of quantization, or changing the quantizer without replacing the in-path compressor, is less direct in these designs. Section~\ref{sec:moti} analyzes how these integration choices affect flexibility, overlap, and error propagation in the GPU-resident workloads targeted by NCCLZ.

To address these challenges, we propose a new integration strategy for compression-enabled collectives for NCCL. The strategy separates quantization from entropy coding and places them at different layers: quantization at the interface layer and entropy coding within NCCL communication primitives. This separation supports flexible composition of compression techniques, isolates the lossless coding stage for compression--communication overlap, and avoids repeated lossy recompression in reduction-heavy collectives. It also preserves NCCL's execution workflow and programming semantics without modifying the transport layer.

%To address these challenges, we propose a new integration strategy for compression-enabled collectives in NCCL. The key idea is to decouple lossy compression into quantization and entropy coding and integrate them into different NCCL layers: quantization is applied at the interface layer, while entropy coding is embedded within NCCL communication primitives. This separation enables flexible composition of compression techniques across different communication scenarios, avoids error propagation in reduction-heavy collectives by integrating only lossless components into the critical communication path, and simplifies overlapping compression with communication by isolating entropy coding overhead. Moreover, the strategy preserves NCCL’s native execution workflow and programming semantics without modifying transport layers, allowing compression to align naturally with NCCL’s optimized GPU-centric design.

Building on this strategy, we develop \sys, a compression-enabled NCCL framework for bandwidth-constrained NVIDIA GPU clusters. \sys combines external, interface-level quantization with entropy coding inside NCCL primitives while preserving NCCL's APIs and execution workflow. As shown in Figure~\ref{fig:arch}, it supports quantizers for scientific and AI workloads and integrates multiple GPU-optimized entropy coders. A lightweight device-side arbitrator selects the most cost-effective runtime path, including bypassing entropy coding, and an overlap-aware mechanism uses NCCL's pipeline to hide coding overhead. We evaluate \sys on scientific datasets, distributed-training gradients, and synthetic workloads across multiple node counts. The results show higher compression ratios and end-to-end throughput than unmodified NCCL and prior compression-assisted CCLs. Our contributions are as follows:

\begin{itemize}
\item We propose a layered integration strategy that separates interface-level quantization from entropy coding inside NCCL primitives, improving flexibility while avoiding repeated lossy recompression.
\item We implement \sys as an opt-in NCCL extension that preserves native APIs, execution order, and grouping semantics.
\item We design a lightweight device-side entropy arbitrator and a two-level overlap mechanism that pipelines coding with NCCL communication.
\item We evaluate \sys on scientific datasets, distributed deep learning workloads, and synthetic benchmarks, observing speedups of up to $9.65\times$ over NCCL and $3.34\times$ over existing compression-assisted CCLs.
\end{itemize}

\section{Background}
\label{bg}
This section summarizes the NCCL execution model and the compression stages relevant to NCCLZ.

\subsection{NCCL}
NCCL (NVIDIA Collective Communications Library) is a widely used GPU communication library for distributed deep learning and other multi-GPU workloads. It is integrated into mainstream frameworks and systems to execute high-throughput collectives~\cite{NCCL,NvidiaNCCLDev,PyTorchDDP,Sergeev2018Horovod,Shoeybi2019MegatronLM,Kwon2023PagedAttention,Zheng2024SGLang}.
NCCL provides collectives such as \texttt{AllReduce}, \texttt{Broadcast}, \texttt{AllToAll}, and \texttt{AllGather}. A collective operates on a group of GPU ranks; for example, \texttt{AllReduce} aggregates values across ranks and returns the result to every rank~\cite{Hu2025DemystifyingNCCL}. Internally, NCCL decomposes collectives into low-level communication \emph{primitives} (e.g., \texttt{send}, \texttt{recv}, and \texttt{recvReduceSend}) that form ring and tree algorithms over different transports and protocols~\cite{Hu2025DemystifyingNCCL}.

An NCCL program creates a communicator for the participating GPU ranks and enqueues collective operations on CUDA streams. To amortize launch overhead, NCCL can group multiple calls and submit them as a batch. Runtime performance then depends on channel-level parallelism, the selected protocol, and topology efficiency~\cite{Hu2025DemystifyingNCCL}.

For large tensors, NCCL partitions data across multiple \emph{communication channels} so disjoint chunks can be processed concurrently. Each channel uses a fixed-size buffer divided into \texttt{NCCL\_STEPS} \emph{slots} for pipelined production, transmission, and consumption~\cite{Hu2025DemystifyingNCCL}. During communicator initialization, NCCL builds logical topologies such as ring and tree and selects a protocol to balance latency and bandwidth, including \texttt{LL}, \texttt{Simple}, and, when supported, \texttt{LL128}~\cite{Hu2025DemystifyingNCCL}.

NCCL uses different data paths for intra-node and inter-node communication~\cite{Hu2025DemystifyingNCCL}.
Within a node, it prioritizes direct GPU peer-to-peer transfers over NVLink or PCIe; the high scale-up bandwidth of NVLink often reduces the benefit of compression~\cite{NVIDIAA100NVLink}.
Across nodes, NCCL uses network transports such as sockets or RDMA, typically with a host-side proxy thread to drive NIC operations for GPU transfers~\cite{Hu2025DemystifyingNCCL}.
In the common baseline path, data is staged through pinned host memory, making performance sensitive to PCIe, CPU progress, and network bandwidth.

\subsection{Lossy Compression}
\label{sec:lossy}
Lossy compression reduces communication volume by encoding numerical tensors into fewer bits at the cost of fidelity loss and extra computation.
In distributed training, error-bounded lossy compression is widely used to improve throughput while preserving convergence through strict distortion guarantees~\cite{Liang2024CommEffSurvey}.
Representative compressors such as SZ~\cite{Di2018PointwiseSZ} and ZFP~\cite{Lindstrom2014ZFP} follow a four-stage pipeline:
(1) \emph{prediction/transform},
(2) \emph{quantization},
(3) \emph{entropy coding}, and
(4) \emph{lossless backend}.
Among them, \emph{quantization} and \emph{entropy coding} largely determine the compression--accuracy trade-off and communication efficiency.

% \textit{Quantization.} As the key mechanism for controlling the trade-off between accuracy and compression ratio, quantization in error-bounded compressors is typically linear and deterministic quantization. Values within a fixed interval are mapped to uniformly spaced bins based on user-specified error bounds. This approach ensures strict error control and predictable reconstruction, making it well-suited for scientific computing and distributed training with numerical guarantees.

% In contrast, QSGD~\cite{Alistarh2017QSGD} adopts stochastic quantization, which uses randomized rounding by assigning each coordinate to a nearby quantization level with probability proportional to its distance from that level. This approach yields unbiased gradient estimates and helps preserve convergence guarantees while significantly reducing communication overhead.

\textbf{\textit{Quantization.}} Quantization maps values to a finite set of levels to reduce representation cost. In error-bounded compressors, it is typically deterministic, assigning values to uniformly spaced bins based on user-specified error bounds and thereby ensuring strict error control. By contrast, QSGD~\cite{Alistarh2017QSGD} uses stochastic quantization, applying randomized rounding with probabilities proportional to distance, yielding unbiased estimates while helping preserve convergence.

% \textit{Entropy coding.}
% Following quantization, entropy coding is applied to further compact the quantized values, often bringing the compressed representation closer to the information-theoretic limit. However, it also introduces additional computational overhead. The choice of entropy coding scheme can significantly impact overall compression efficiency under different data distributions.

% Fixed-length coding, as used in cuSZp~\cite{Huang2023CuSZp}, assigns the same number of bits to each quantized symbol, regardless of its frequency. This approach minimizes encoding and decoding complexity and is highly amenable to parallelization, making it particularly suitable for GPU-based implementations. It performs well when the symbol distribution is close to uniform, where variable-length coding offers limited advantage.

% In contrast, Huffman coding, employed in compressors like SZ~\cite{Di2018PointwiseSZ}, assigns shorter codes to more frequent symbols based on their probability. This enables better compression in the presence of skewed distributions but requires additional overhead for codebook construction and can be less efficient on parallel hardware due to its sequential nature.

\textbf{\textit{Entropy coding.}} Entropy coding encodes quantized symbols more compactly according to their statistical distribution, at the cost of extra computation. Its coding efficiency can vary significantly across different data distributions. Fixed-length coding, as used in cuSZp~\cite{Huang2023CuSZp}, assigns the same number of bits to each symbol, minimizing coding complexity and favoring parallel execution, especially on GPUs. It is effective when the symbol distribution is close to uniform. By contrast, Huffman coding, used in compressors such as SZ~\cite{Di2018PointwiseSZ}, assigns shorter codes to more frequent symbols, achieving better compression for skewed distributions at the cost of codebook overhead and lower parallel efficiency.

\section{Problem Analysis and Design Motivation}
\label{sec:moti}
%New paragraph: Existing compression-enabled CCLs mainly adopt two integration strategies. \emph{MPI-based} approaches extend CUDA-aware MPI stacks by embedding compression into the communication path or collective workflow, typically treating a complete compressor as part of collective execution so tensors are compressed before transfer and decompressed when communication or reduction consumes them~\cite{ZhouCKKGS021,Huang2024gZCCL,Huang2025ghZCCL}. In contrast, \emph{GPU-aware} approaches such as COCCL build compression-aware workflows around a vendor CCL by re-wrapping GPU-native collective APIs and injecting only lightweight device-side \emph{quantization} into the collective path, so the reduced-precision payload can still follow NCCL's native GPU data-transfer path~\cite{COCCLRepo}. However, this strategy mainly relies on quantization itself for byte reduction rather than integrating a full quantization-plus-entropy-coding pipeline into the collective runtime; in particular, entropy coding is not deeply leveraged inside the communication engine. 
The placement of compression stages within a CCL determines which components can be selected independently and which costs can overlap with communication. COCCL~\cite{Liu2026COCCL} re-wraps NCCL calls at the API layer and overlaps lightweight, LLM-oriented quantization with communication on multiple CUDA streams~\cite{COCCLRepo}. MPI-based frameworks such as MVAPICH2~\cite{ZhouCKKGS021}, gZCCL~\cite{Huang2024gZCCL}, and ghZCCL~\cite{Huang2025ghZCCL} instead integrate complete scientific compressors into an MPI communication path and overlap compressor execution with MPI primitives. Both approaches are effective in their target settings, but their stage boundaries impose trade-offs when scientific and AI quantizers must coexist with independently selectable entropy coding.

%Existing compression-enabled CCLs mainly follow two integration approaches. \emph{MPI-based} approaches, such as MVAPICH2, gZCCL, and ghZCCL~\cite{ZhouCKKGS021,Huang2024gZCCL,Huang2025ghZCCL}, integrate complete compressors into the MPI communication path or collective workflow. In these designs, quantization, entropy coding, and associated metadata handling are treated as part of collective execution, and overlap is mainly achieved through collective-level pipelining or communication--compression co-scheduling to hide the cost of compression. On the other hand, \emph{GPU-aware} approaches such as COCCL adopt only lightweight quantization-based compression (SDP4Bit and min-max uint8) and place it at the NCCL API layer by re-wrapping NCCL calls~\cite{COCCLRepo}. This design mainly relies on the low overhead of quantization itself, and overlaps compression and communication through multistreaming scheme.

\textbf{\textit{Limited stage-level flexibility.}} COCCL supports multiple quantization methods for LLMs (e.g., \texttt{SDP4Bit} and \texttt{minmaxUint8})~\cite{COCCLRepo}, but it does not expose entropy coding as an independent stage. Conversely, the cited MPI-based approaches select or integrate complete scientific compressors. In both cases, changing the quantizer independently of the entropy coder, or enabling and disabling entropy coding without replacing the full compressor path, requires additional integration work. NCCLZ targets this stage-level flexibility rather than claiming that the same separation is infeasible in MPI.

\textbf{\textit{Overlap granularity.}} Entropy coding is often the most expensive lossless stage and is therefore important to overlap with communication. Existing systems overlap quantization or a complete compressor pipeline with collective progress. Those strategies demonstrate that overlap is practical in MPI, but they do not directly use NCCL's device-side primitive and slot pipeline. Because NCCL's GPU-native path can leave a different and sometimes shorter overlap window, NCCLZ places only entropy coding inside this path and retains quantization at the interface.

\textbf{\textit{Error propagation in reduction collectives.}} When lossy quantization is placed inside a reduction-heavy collective such as \texttt{AllReduce}, intermediate results may be repeatedly decompressed and recompressed~\cite{mvapich2gdr_userguide,Huang2024gZCCL}. The resulting dequantization and requantization can accumulate distortion. ghZCCL avoids this behavior through homomorphic compression, although the applicability and cost of that compressor depend on the workload. NCCLZ instead keeps quantization outside the collective path and integrates only lossless entropy coding into NCCL primitives.

These trade-offs motivate an integration strategy that separates scientific or AI quantization from on-demand entropy coding, aligns the lossless stage with NCCL's pipeline, and avoids repeated lossy recompression (Sec.~\ref{sec:strategy_design}).

\section{\sys Design and Implementation}
\label{ncclz}
In this section, we present the design of \sys, an opt-in compression-aware extension to NCCL that targets bandwidth-dominated inter-node communication.

\subsection{Integration Strategy Design}
\label{sec:strategy_design}
We separate compression into two stages and integrate them at different layers. Quantization is performed once at the interface boundary, before NCCL communication primitives are invoked, while lossless entropy coding is integrated into those primitives and applied only when it is expected to reduce end-to-end time. This separation makes the approximation boundary explicit, keeps lossy numerical semantics outside the collective runtime, and aligns the in-runtime lossless stage with NCCL's data path.

This addresses the trade-offs identified in Sec.~\ref{sec:moti}. The interface selects an application quantizer while the runtime independently selects an entropy-coding path. Placing only the lossless stage inside NCCL reduces the in-path work that must be overlapped while preserving native execution and grouping semantics. Keeping quantization outside NCCL primitives also prevents repeated lossy recompression of intermediate reduction results.

\subsection{\sys Overview}
\begin{figure*}[t]
    \centering
    \includegraphics[width=1.01\textwidth]{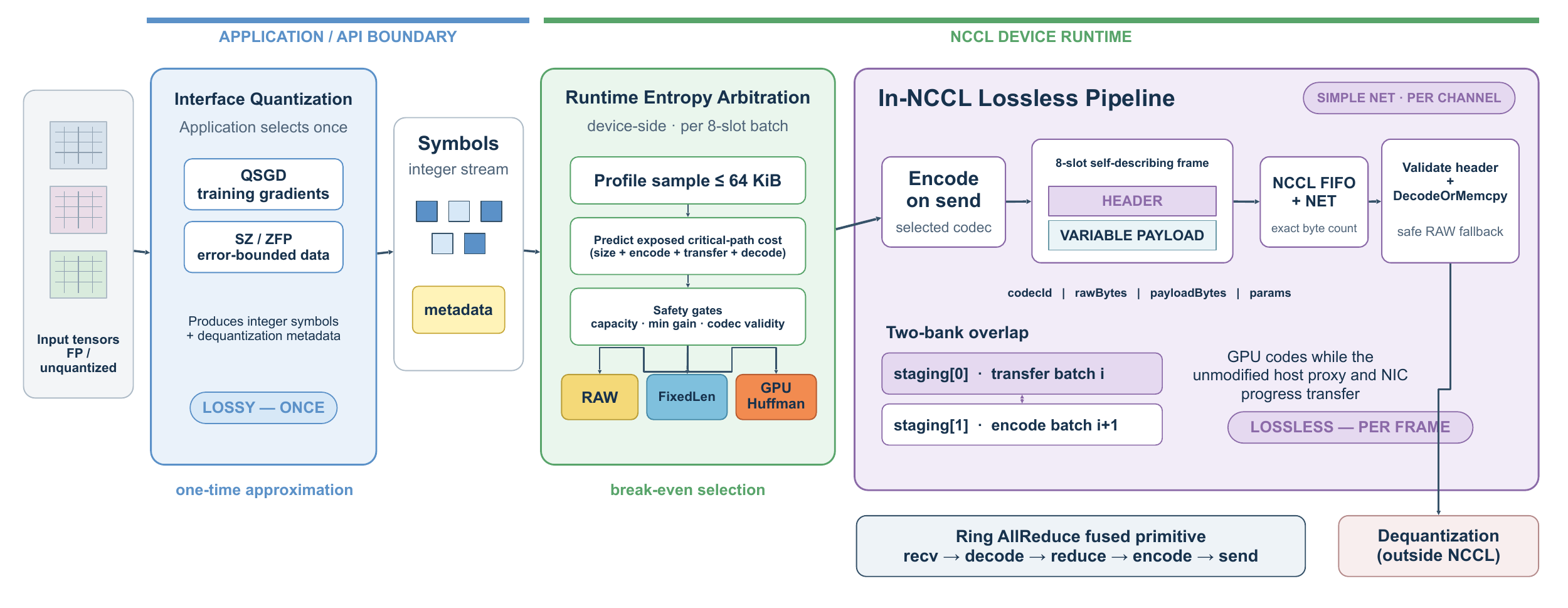}
    \caption{Layered NCCLZ design. The application selects interface-level quantization, device-side REA selects \textsc{Raw}, \textsc{FixedLen}, or GPU \textsc{Huffman}, and NCCL primitives pipeline coding with NET communication.}
    \vspace{-9pt}
    \label{fig:nccldesign}
\end{figure*}

Figure~\ref{fig:nccldesign} shows the three components of \sys: an external quantizer that applies the one-time approximation before data enter NCCL, a device-side Runtime Entropy Arbitration (REA) module that selects among \textsc{Raw}, \textsc{FixedLen}, and GPU \textsc{Huffman} paths (Sec.~\ref{sec:ncclz-selector}), and an in-NCCL mechanism that pipelines encoding and decoding with communication and fused reduction progress (Sec.~\ref{sec-ncclz-overlap}).

For unquantized input, the external quantizer produces an integer symbol stream before communication. The application explicitly selects the mode through the API or configuration; \sys does not infer it from tensor values. We use QSGD-style stochastic quantization for gradients~\cite{Alistarh2017QSGD} and deterministic error-bounded quantization for other numerical inputs, such as SZ/ZFP-style bounded quantization~\cite{Di2018PointwiseSZ,Lindstrom2014ZFP}. The quantizer also produces the metadata required for dequantization, such as scale parameters, which is stored in the batch header or a small companion buffer sent with the payload. Because the subsequent entropy-coding stage is lossless, quantization is the only source of approximation.

REA examines a sample of up to 64\,KiB and uses constant-time gates, including the size threshold and Huffman-context validity, to decide whether to evaluate fixed-length packing, GPU Huffman coding, or both. It estimates the expected payload size, applies a minimum-gain threshold, and performs strict capacity checks. On success, the selected path materializes a self-describing frame with Header + Payload; the header records \texttt{codecId}, \texttt{rawBytes}, \texttt{payloadBytes}, and codec-specific parameters. When compression is not beneficial or staging space is insufficient, \sys safely falls back to \textsc{Raw}. REA, sample profiling, codec selection, encoding, and decoding execute in GPU device code; the host-side NCCL proxy only drives NIC progress and transfers the byte count published through existing FIFO metadata.

The in-NCCL integration attaches encode-on-send and decode-on-receive operations at primitive boundaries, so compressed frames retain NCCL's ordering and progress semantics. \sys fixes the framing granularity to an 8-slot batch: each frame spans exactly eight slots, with a raw upper bound of approximately \texttt{NCCL\_BUFFSIZE} (4\,MiB by default), and lays the compressed payload contiguously across them to form a larger, stable coding window. Two ping-pong staging banks, \texttt{staging[0..1]}, alternate across batches, allowing batch $i{+}1$ to be encoded while batch $i$ is in flight. The receiver validates and decodes each frame independently, with a robust \textsc{Raw} fallback when validation or decoding fails. For \texttt{AllReduce}, \sys uses the fused \texttt{recvReduceSend} primitive to stream \texttt{recv$\rightarrow$decode$\rightarrow$reduce$\rightarrow$encode$\rightarrow$send}, overlapping neighboring chunks and hiding coding overhead behind the reduction pipeline.

\subsection{Runtime Entropy Arbitration}
\label{sec:ncclz-selector}
The device-side \emph{runtime entropy arbitration} (REA) module chooses whether to send a quantized symbol stream as \textsc{Raw}, \textsc{FixedLen}, or GPU \textsc{Huffman}. Its decision depends on message size, sampled compressibility, and the transport regime. REA treats codec selection as a runtime break-even decision: it compares predicted critical-path costs rather than ranking candidates by compressed size alone.

\textbf{Arbitration mechanism.}
\emph{When \textsc{Raw} is preferred.}
REA selects \textsc{Raw} when high effective bandwidth or a small message makes the exposed encode/decode overhead larger than the saved transfer time. This commonly occurs on intra-node paths such as NVLink or other direct GPU-resident P2P transports, where transmission contributes less to end-to-end time while codec execution remains on the critical path. Thus, \textsc{Raw} is the optimal high-bandwidth decision rather than merely a fallback after failed compression.

\emph{When entropy coding is preferred.}
Entropy coding is preferred for the bandwidth-dominated inter-node regime targeted by \sys. REA profiles a bounded sample to estimate each codec's payload reduction. For \textsc{FixedLen}, the estimate follows the effective symbol width implied by the sample; for GPU \textsc{Huffman}, it follows the sampled distribution skew under the cached codebook. \textsc{FixedLen} is favored when the symbol range is narrow and low-overhead packing is sufficient, whereas \textsc{Huffman} is selected only when the greater reduction from additional skew can amortize its extra codec cost. Huffman also requires a valid context and a minimum message size.

Let $B$ denote the raw payload size, $\widehat{P}_c$ the predicted transmitted size under codec $c$, and $\widehat{\beta}_{\mathrm{eff}}(\pi)$ the calibrated effective bandwidth for transport hint $\pi$. REA estimates the exposed critical-path time as
{\setlength{\abovedisplayskip}{4pt}
\setlength{\belowdisplayskip}{4pt}
\setlength{\abovedisplayshortskip}{2pt}
\setlength{\belowdisplayshortskip}{2pt}
\begin{equation}
\label{eq:rea-time-compact}
\widehat{T}_c
=
\alpha_c
+
\lambda_c^{\mathrm{enc}}\widehat{E}_c
+
\frac{\widehat{P}_c}{\widehat{\beta}_{\mathrm{eff}}(\pi)}
+
\lambda_c^{\mathrm{dec}}\widehat{D}_c,
\end{equation}
}
where $\widehat{E}_c$ and $\widehat{D}_c$ denote the estimated encode and decode costs, and $\lambda_c^{\mathrm{enc}},\lambda_c^{\mathrm{dec}}\in[0,1]$ represent the fractions that remain exposed after overlap. For \textsc{Raw}, $\widehat{P}_{\mathsf{R}}=B$ and codec costs are negligible; for entropy codecs, $\widehat{P}_c$ is predicted from the bounded sample together with codec-specific metadata overheads.

REA admits an entropy candidate only if its predicted payload fits the staging capacity, its predicted compression gain exceeds the minimum-gain threshold, and its codec-specific enable conditions hold. It then selects
\begin{equation}
\label{eq:rea-select-compact}
c^\star
=
\arg\min_{c\in \{\mathsf{R}\}\cup\mathcal{C}_{\mathrm{adm}}}
\widehat{T}_c,
\end{equation}
where $\mathcal{C}_{\mathrm{adm}}$ is the set of admissible entropy codecs. If no entropy codec survives these checks, or if the realized output after materialization fails capacity or gain validation, REA commits a safe \textsc{Raw} frame. This makes the arbitration boundary explicit while keeping the mathematical model subordinate to runtime decision logic.

\textbf{Arbitration procedure.}
Algorithm~\ref{alg:ncclz-selector-compact} implements this decision in four stages. Its inputs are the raw stream $\textit{raw}[0..\textit{rawBytes})$, staging destination $\textit{outHdrBase}$ with capacity $\textit{stageCapBytes}$, pre-initialized Huffman context $\mathsf{huffCtx}$, Huffman threshold $T_{\text{huff}}$, and minimum-gain requirement $\textit{minGainPermil}$.

Lines~1--4 reject degenerate inputs and derive the payload budget; Lines~5--7 directly commit a \textsc{Raw} frame when arbitration is unnecessary. Otherwise, Lines~8--11 obtain a transport hint, profile a bounded sample, and call \Call{ArbitratePlan}{} using the message size, compressibility estimate, transport regime, and codec constraints. Lines~12--17 materialize a selected non-\textsc{Raw} plan and accept it only after \Call{AcceptRealized}{} validates the output. Line~18 calls \Call{CommitRawOrFail}{}, which commits \textsc{Raw} if it fits or returns failure. The procedure therefore preserves explicit early guards, planning, selective materialization, post-validation, and safe fallback.

\begin{algorithm}[t]
\caption{Device-side runtime entropy-arbitration procedure.}
\label{alg:ncclz-selector-compact}
\scriptsize
\begin{algorithmic}[1]
\Statex \textbf{Input:} raw[0..rawBytes), outHdrBase, stageCapBytes, huffCtx, $T_{\mathrm{huff}}$, minGainPermil
\Statex \textbf{Output:} return $(codecId, payloadBytes, totalBytes)$ and write a frame at outHdrBase on success
\If{rawBytes $\le 0$ \textbf{or} stageCapBytes $\le$ HdrBytes}
  \State \Return (0,0,0)
\EndIf
\State payloadCap $\gets$ stageCapBytes $-$ HdrBytes
\If{\Call{SmallBatchFastPath}{rawBytes}}
  \State \Return \Call{CommitRawOrFail}{raw, rawBytes, outHdrBase, payloadCap}
\EndIf
\State hint $\gets$ \Call{TransportHint}{}
\State sampleBytes $\gets$ \Call{SampleWindow}{rawBytes}
\State stats $\gets$ \Call{ProfileSample}{raw[0:sampleBytes], huffCtx}
\State plan $\gets$ \Call{ArbitratePlan}{rawBytes, payloadCap, stats, hint, huffCtx, $T_{\mathrm{huff}}$, minGainPermil}
\If{plan.codecId $\neq$ \textsc{Raw}}
  \State $P \gets$ \Call{EncodePlan}{plan, raw, rawBytes, outHdrBase, payloadCap, huffCtx}
  \If{\Call{AcceptRealized}{plan.codecId, $P$, rawBytes, payloadCap, minGainPermil}}
    \State \Return (plan.codecId, $P$, HdrBytes+$P$)
  \EndIf
\EndIf
\State \Return \Call{CommitRawOrFail}{raw, rawBytes, outHdrBase, payloadCap}
\end{algorithmic}
\end{algorithm}

\textbf{Arbitration overhead.}
Constant-time gates check the small-batch path, Huffman threshold, and context validity. Sampling touches at most 64~KiB per 8-slot batch, and one control thread produces one estimate without scanning the full payload. Planning uses a max-$|x|$ scan for \textsc{FixedLen} or a 256-bin sample histogram and expected-code-length calculation for \textsc{Huffman}; only the selected codec processes the full frame.

\subsection{Overlapping Entropy Coding with Communication}
\label{sec:ncclz-integrated}

\sys overlaps only the lossless entropy-coding stage within NCCL's per-channel slot pipeline. MPI-based systems instead overlap a complete compressor or collective-level compression pipeline with MPI progress~\cite{Zhou2021MVAPICHCompression,Huang2024gZCCL,Huang2025ghZCCL}; COCCL overlaps lightweight quantization through re-wrapped NCCL APIs, whose compression-supported calls cannot be enclosed by \texttt{ncclGroupStart}/\texttt{ncclGroupEnd}~\cite{COCCLRepo}. \sys integrates at device-side primitive boundaries so coding can proceed alongside NET transfers without changing native ordering or grouping semantics.

\subsubsection{Primitive Coverage and Full-Duplex Collectives}
\sys places framing and codec selection at the communication boundaries of NCCL device primitives~\cite{Hu2025DemystifyingNCCL}. It encodes immediately before an outgoing payload is enqueued in the connection FIFO and decodes immediately after an incoming frame is dequeued, before subsequent copy or reduction logic. The same boundary covers primitives with concurrent send and receive operations, such as \texttt{recvReduceSend} in \texttt{AllReduce}, without specializing for a particular collective algorithm.

\subsubsection{Batch Framing (Header + Payload)}
Each entropy-coded unit is a self-describing frame with a compact header and variable-length payload. The header records \texttt{codecId}, \texttt{rawBytes}, \texttt{payloadBytes}, and codec-specific parameters, allowing the receiver to validate and decode each frame independently or safely fall back to a raw copy.

\begin{figure}[!t]
  \centering
  \includegraphics[width=0.88\linewidth,height=0.7\linewidth]{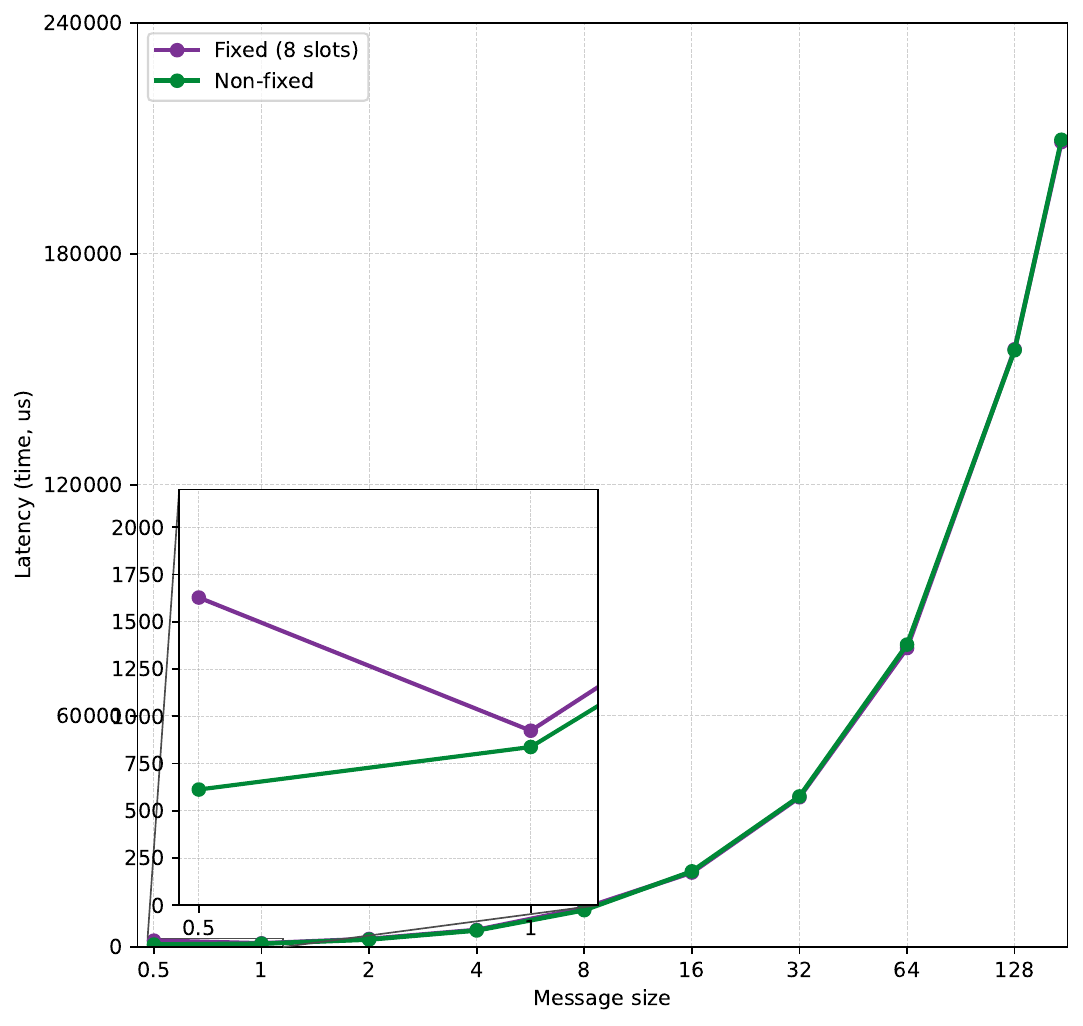}
  \vspace{-8pt}
  \caption{Latency of fixed 8-slot batching and non-fixed NCCL batching for message sizes from 0.5 to 128\,MiB. The inset enlarges the 0.5- and 1-MiB results.}
    \vspace{-9pt}

  \label{fig:slot-batching}
\end{figure}

\begin{figure*}[!b]
    \centering
    \includegraphics[width=0.85\textwidth]{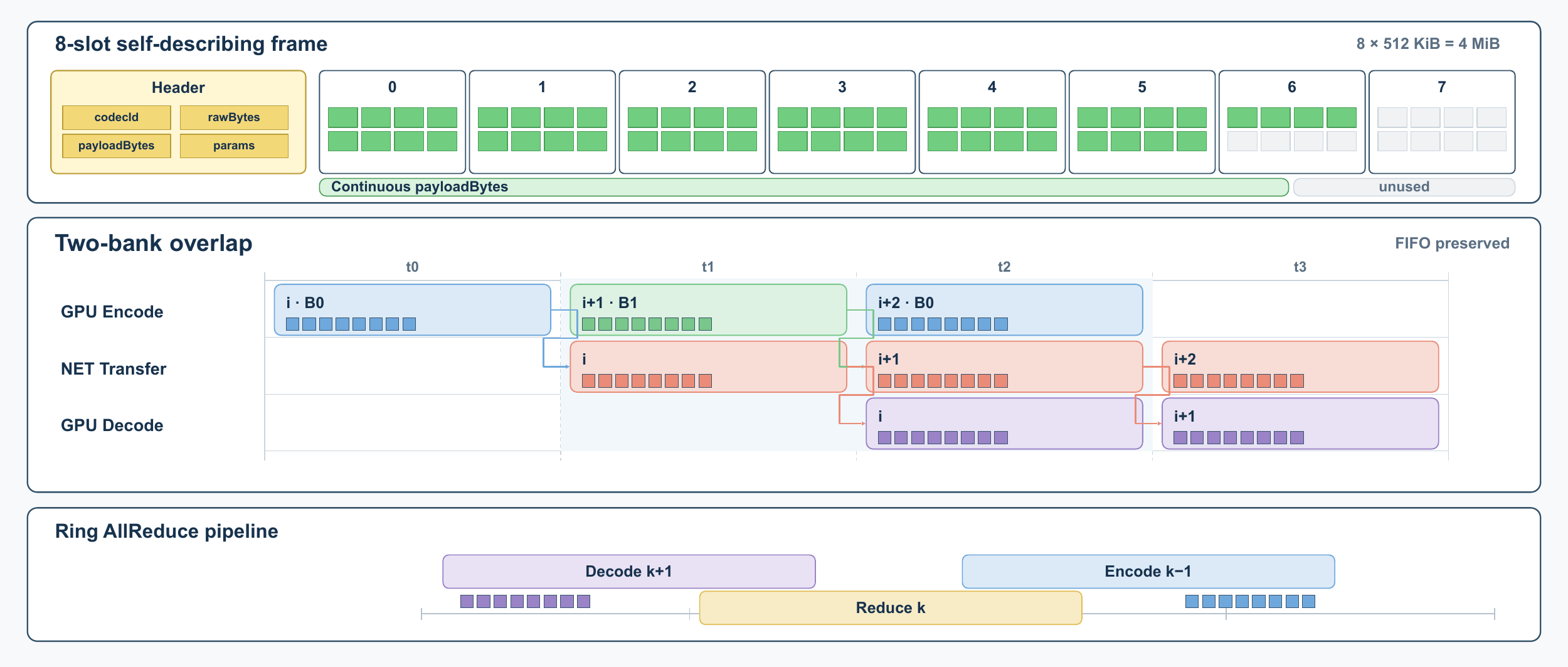}
    \caption{Two-bank, batch-level overlap. While one 8-slot batch is transferred, the alternate staging bank is used to encode or decode an adjacent batch.}
    \vspace{-9pt}
    \label{fig:firstlayer}
\end{figure*}

\subsubsection{8-slot batching}
\label{sec:ncclz-8slot}
NCCL Simple uses a fixed per-channel buffer of \texttt{NCCL\_BUFFSIZE}=4\,MiB, partitioned into \texttt{NCCL\_STEPS}=8 slots of 512\,KiB each~\cite{Hu2025DemystifyingNCCL}. \sys uses one complete 8-slot channel buffer as the framing and coding unit.

Entropy coding produces variable-length output, whereas NCCL Simple uses fixed-slot FIFO bookkeeping between the GPU kernel and host proxy. Padding every compressed slot would waste bandwidth, while changing slot accounting would violate proxy expectations. The 8-slot frame instead publishes a byte-accurate length, so the proxy transfers only valid bytes while preserving slot rotation and FIFO order. It also provides up to 4\,MiB of coding context, amortizing selector and header overhead.

Figure~\ref{fig:slot-batching} reports message size in MiB. Small messages exhibit fill-and-drain effects, but fixed and non-fixed batching converge after link saturation. Thus, fixed 8-slot framing preserves performance in the target bandwidth-bound regime while increasing coding context and amortizing per-frame overhead.

\subsubsection{Two-Level Overlap}
\label{sec-ncclz-overlap}

Entropy coding adds GPU work to the critical path unless it overlaps an existing bottleneck. \sys targets the bandwidth-dominated Simple protocol on the inter-node NET path, where the NIC and NCCL host proxy progress network transfers while the GPU kernel may wait for FIFO slots and steps. Encoding and decoding can use part of these wait intervals at both batch and primitive granularity.

\paragraph{Batch-Level Overlap via Ping-Pong Staging}
Two per-channel staging banks, \texttt{staging[0..1]}, alternate across 8-slot batches. After batch $i$ is published to the connection FIFO, the GPU can encode batch $i{+}1$ in the other bank while the NET proxy and NIC transfer batch $i$. This avoids overwriting in-flight output and preserves FIFO order because publication still follows NCCL's slot-availability protocol. On the receive side, \sys decodes a dequeued batch while the proxy and NIC transfer the next frame. This cross-batch overlap relies only on existing NCCL progress mechanisms.

\paragraph{In-Primitive Overlap for Ring \texttt{AllReduce}}
For ring \texttt{AllReduce}, fused primitives such as \texttt{recvReduceSend} form a streaming dependency chain. \sys decodes immediately after dequeue and encodes immediately before enqueue, producing the per-chunk pipeline \texttt{receive $\rightarrow$ decode $\rightarrow$ reduce $\rightarrow$ encode $\rightarrow$ send}. Reduction of chunk $k$ can overlap decoding of $k{+}1$ and encoding of $k{-}1$. This staggered execution reduces bubbles without changing NCCL's collective schedule; REA selects \textsc{Raw} when a message cannot amortize the codec work.

\begin{algorithm}[t]
\caption{NCCLZ entropy-coding pipeline for the NET path.}
\label{alg:ncclz-pipeline}
\scriptsize
\begin{algorithmic}[1]
\Statex \textbf{Input:} connection state $\mathsf{conn}$ (FIFO, 8 slots), two ping-pong staging banks $\mathsf{staging}[0..1]$,
\Statex \hspace{2.2em} sender buffer $\textit{raw}[0..\textit{rawBytes})$, receiver buffer $\textit{dst}[0..\textit{rawBytes})$,
\Statex \hspace{2.2em} Huffman context $\mathsf{huffCtx}$, $T_{\text{huff}}$, $\textit{minGainPermil}$
\Statex \textbf{Output:} sender enqueues a self-describing frame;
\Statex \hspace{2.2em} receiver reconstructs $\textit{dst}$ (lossless) with robust fallback

\Statex \textbf{Sender-side (per 8-slot batch):}
\State $\mathsf{bank} \leftarrow \mathsf{batchId} \bmod 2$
\State \Call{WaitSlotsFree}{$\mathsf{conn}, 8$}
\State $\textit{outHdrBase} \leftarrow \mathsf{staging}[\mathsf{bank}]$
\State $\textit{stageCapBytes} \leftarrow \Call{StageCapBytes}{\mathsf{staging}[\mathsf{bank}]}$
\State $(\mathsf{codecId}, \textit{payloadBytes}, \textit{totalBytes}) \leftarrow$
\Statex \hspace{1.4em}$\Call{encode\_best}{\textit{raw},\textit{rawBytes},\ldots,\textit{minGainPermil}}$
\If{$\textit{totalBytes}>0$}
  \State \Call{EnqueueFrameToFifo8}{$\mathsf{conn}, \textit{outHdrBase}, \textit{totalBytes}$}
\EndIf

\Statex \textbf{Receiver-side (per 8-slot batch):}
\State $\textit{frameBase} \leftarrow \Call{DequeueFrameFromFifo8}{\mathsf{conn}}$
\State $\mathsf{hdr} \leftarrow \Call{ParseHeader}{\textit{frameBase}}$
\If{\textbf{not} \Call{ValidateHeader}{$\mathsf{hdr}$}}
  \State \Call{Memcpy}{$\textit{dst}, \textit{frameBase}+\textit{HdrBytes}, \textit{rawBytes}$}
\Else
  \State $\mathsf{codecId} \leftarrow \mathsf{hdr.codecId}$;\ \ $\textit{payloadBytes} \leftarrow \mathsf{hdr.payloadBytes}$
  \State \Call{DecodeOrMemcpy}{$\mathsf{codecId}, \textit{frameBase}+\textit{HdrBytes}, \textit{payloadBytes}, \textit{dst}, \mathsf{hdr.rawBytes}, \mathsf{huffCtx}, \mathsf{hdr.params}$}
\EndIf
\end{algorithmic}
\end{algorithm}

\subsection{\sys Implementation}
\label{sec:ncclz-impl}
All components are integrated into NCCL as an opt-in extension. The two codecs and REA are header-only CUDA modules (\texttt{.cuh}) invoked by a small dispatcher in NCCL device code, minimizing changes to the build and link process. On the NET path of the Simple protocol, \sys intercepts each 8-slot batch immediately before FIFO enqueue and after FIFO dequeue. The GPU performs arbitration and coding, while the unmodified host proxy consumes the published frame length and drives NIC progress.

Algorithm~\ref{alg:ncclz-pipeline} summarizes the workflow. Its inputs are connection state $\mathsf{conn}$, ping-pong banks $\mathsf{staging}[0..1]$, user buffers $\textit{raw}$ and $\textit{dst}$, and codec controls $\mathsf{huffCtx}$, $T_{\text{huff}}$, and $\textit{minGainPermil}$. The sender produces a self-describing frame, and the receiver reconstructs $\textit{dst}$ losslessly, with a safe \textsc{Raw} fallback if header validation fails.

\paragraph{Sender Side (Lines~1--8)}
The sender selects a staging bank using $\mathsf{batchId}\bmod 2$ and waits for eight free FIFO slots (Lines~1--2), then sets the frame base and capacity (Lines~3--4). The \texttt{encode\_best} call selects \textsc{FixedLen}, \textsc{Huffman}, or \textsc{Raw} and materializes the complete frame in place (Line~5). The sender publishes it only if $\textit{totalBytes}>0$ (Lines~6--8), preventing partial-frame exposure.

\paragraph{Receiver Side (Lines~9--16)}
The receiver dequeues the next 8-slot frame and parses its header (Lines~9--10). If validation fails, it uses \textsc{Raw} semantics and copies the expected bytes into $\textit{dst}$ (Lines~11--12). Otherwise, it reads $\textit{codecId}$ and $\textit{payloadBytes}$ and calls \texttt{DecodeOrMemcpy} to decode \textsc{FixedLen} or \textsc{Huffman}, or copy a \textsc{Raw} payload (Lines~13--16).

% requires:
% \usepackage{booktabs}
% \usepackage{multirow}
% \usepackage{adjustbox}

\begin{table*}[t]
\centering
\footnotesize
\setlength{\tabcolsep}{4pt}
\renewcommand{\arraystretch}{1.05}

\caption{Compression ratios (CRs) on 2, 4, and 8 nodes for scientific datasets (REL=$10^{-4}$) and training gradients. Quant-only applies no entropy coding; the other columns apply \textsc{FixedLen} or GPU \textsc{Huffman} after quantization.}
\label{tab:cr_twocol_quant_fixed_huff_no_overall}

\begin{adjustbox}{max width=\textwidth}
\begin{tabular}{@{}lcccccccccccc@{}}
\toprule
\multirow{2}{*}{\textbf{Dataset}} &
\multicolumn{3}{c}{\textbf{Quant-only (CR)}} & \textbf{Avg} &
\multicolumn{3}{c}{\textbf{FixedLen (CR)}}   & \textbf{Avg} &
\multicolumn{3}{c}{\textbf{GPU Huffman (CR)}}& \textbf{Avg} \\
\cmidrule(lr){2-4}\cmidrule(lr){6-8}\cmidrule(lr){10-12}
& \textbf{2-node} & \textbf{4-node} & \textbf{8-node} & \textbf{(2/4/8)} &
  \textbf{2-node} & \textbf{4-node} & \textbf{8-node} & \textbf{(2/4/8)} &
  \textbf{2-node} & \textbf{4-node} & \textbf{8-node} & \textbf{(2/4/8)} \\
\midrule
QMCPack (1.2\,GiB) &
3.05$\times$ & 3.10$\times$ & 3.00$\times$ & \textbf{3.05$\times$} &
4.61$\times$ & 4.59$\times$ & 4.68$\times$ & \textbf{4.63$\times$} &
8.15$\times$ & 8.25$\times$ & 8.35$\times$ & \textbf{8.25$\times$} \\
CESM-ATM (2.0\,GiB) &
3.20$\times$ & 3.25$\times$ & 3.18$\times$ & \textbf{3.21$\times$} &
6.74$\times$ & 7.24$\times$ & 6.35$\times$ & \textbf{6.78$\times$} &
9.45$\times$ & 9.65$\times$ & 9.85$\times$ & \textbf{9.65$\times$} \\
Grad (QSGD, 150\,MiB) &
2.55$\times$ & 2.60$\times$ & 2.52$\times$ & \textbf{2.56$\times$} &
5.26$\times$ & 6.14$\times$ & 6.45$\times$ & \textbf{5.95$\times$} &
8.44$\times$ & 7.31$\times$ & 8.47$\times$ & \textbf{8.07$\times$} \\
\bottomrule
\end{tabular}
\end{adjustbox}
\end{table*}

\section{Evaluation}
This section describes the experimental methodology and analyzes the results.

\subsection{Experimental setup}
\paragraph{Platform and software}
We conduct all experiments on the Polaris supercomputer at Argonne National Laboratory~\cite{alcf-polaris}.
Each Polaris compute node has an AMD EPYC Milan CPU, 512\,GB of system memory, and four NVIDIA A100 GPUs. The nodes are connected by an HPE Slingshot network.
Across all experiments, we use unmodified NCCL 2.28.3 as the baseline and keep the software stack and runtime configuration consistent, including the same \texttt{NCCL\_PROTO}, \texttt{NCCL\_ALGO}, and other relevant environment variables.

\paragraph{Datasets}
We use two scientific datasets from SDRBench, QMCPack and CESM-ATM, and collect per-epoch gradients from ResNet-18 and ResNet-34 to represent distributed-training traffic. To isolate scalability effects across node counts and message sizes, we also use the synthetic FP32 payloads provided by \texttt{nccl-tests}.

\subsection{Compression Ratio Study}
\label{sec:cr-study}

We first measure NCCLZ's compression ratio on the scientific datasets and training gradients. We apply deterministic, error-bounded quantization with a relative error bound of REL=$10^{-4}$ to QMCPack and CESM-ATM, and apply QSGD to one epoch of training gradients. We run \texttt{AllReduce} on 2, 4, and 8 nodes. To isolate the two entropy coders, we pin NCCLZ to \textsc{FixedLen} or \textsc{GPU Huffman} and report the resulting ratios in Table~\ref{tab:cr_twocol_quant_fixed_huff_no_overall}.

We use the raw, pre-quantization byte count as the reference: $\mathrm{CR}=B_{\text{raw}}/B_{\text{out}}$. The quantization-only ratio is $\mathrm{CR}_{\text{quant}}=B_{\text{raw}}/B_{\text{quant}}$, where $B_{\text{quant}}$ is the payload size after quantization but before entropy coding. The end-to-end ratio is $\mathrm{CR}_{\text{final}}=B_{\text{raw}}/B_{\text{entropy}}$, where $B_{\text{entropy}}$ is the final entropy-coded payload size.

Across all datasets and node counts, \textsc{GPU Huffman}
achieves a higher CR than \textsc{FixedLen}, suggesting that the
quantized symbol streams retain distribution skew that Huffman
coding can exploit. In contrast, \textsc{FixedLen} primarily removes
representational slack through bit-width trimming. Averaged across
node counts, QMCPack increases from 3.05$\times$ with quantization
alone to 4.63$\times$ with \textsc{FixedLen} and 8.25$\times$ with
\textsc{GPU Huffman}. The gradient tensor increases from
2.56$\times$ to 5.95$\times$ and 8.07$\times$, respectively.
Entropy coding therefore provides a second reduction stage after
quantization, reducing the communicated payload without changing
the quantized reconstruction.

\begin{figure}[b]
  \centering
  \includegraphics[width=0.99\linewidth]{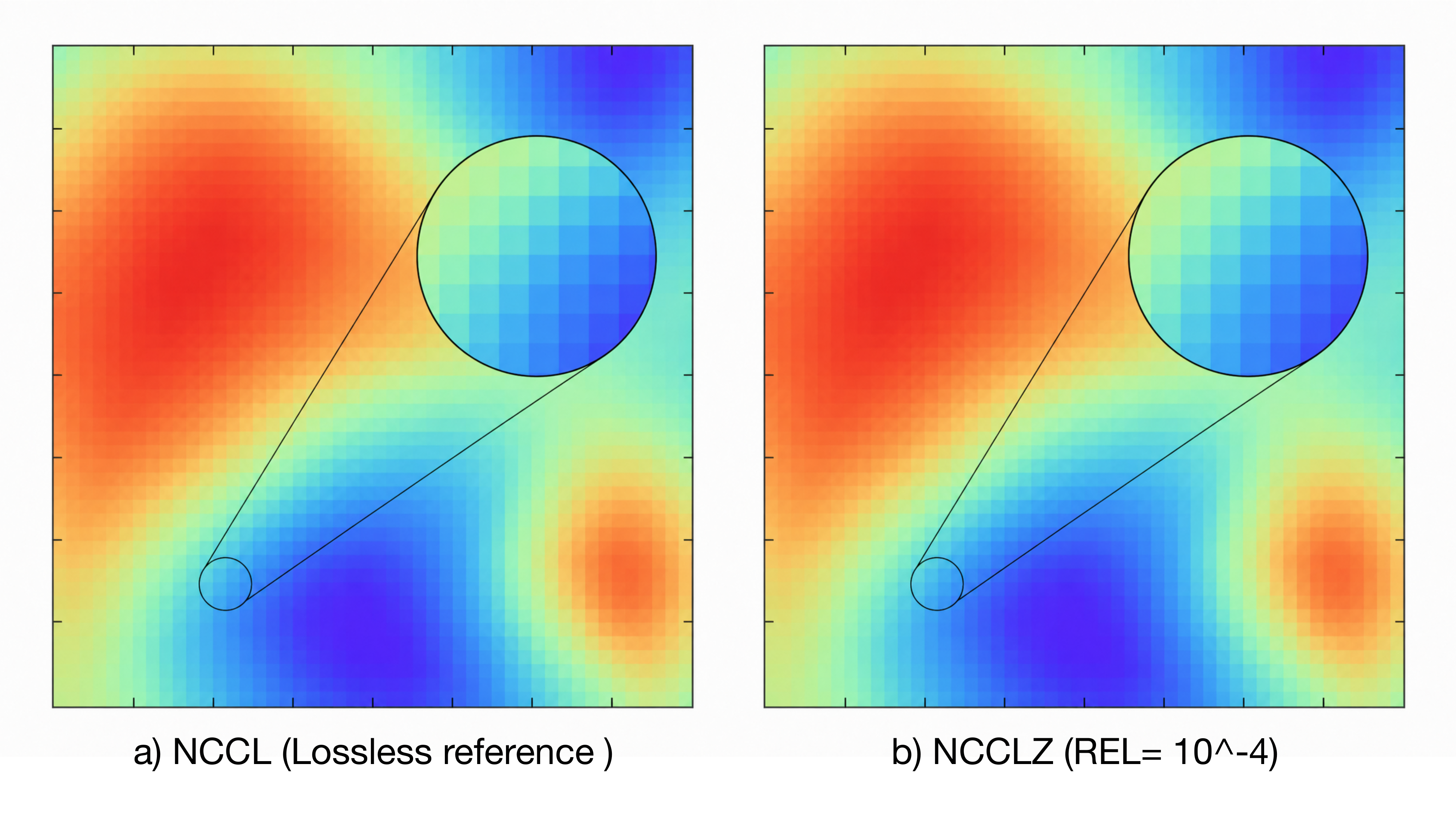}
  \caption{Visual quality comparison on QMCPack (the 700th slice
  along the third dimension) between (a) unmodified NCCL, which
  serves as the lossless reference, and (b) NCCLZ with
  REL=$10^{-4}$. Both panels use identical color limits, and the
  circular insets magnify the same spatial region.}
  \label{fig:qmcpack}
  \vspace{-8pt}
\end{figure}

\begin{figure*}
    \centering
    \includegraphics[width=\linewidth]{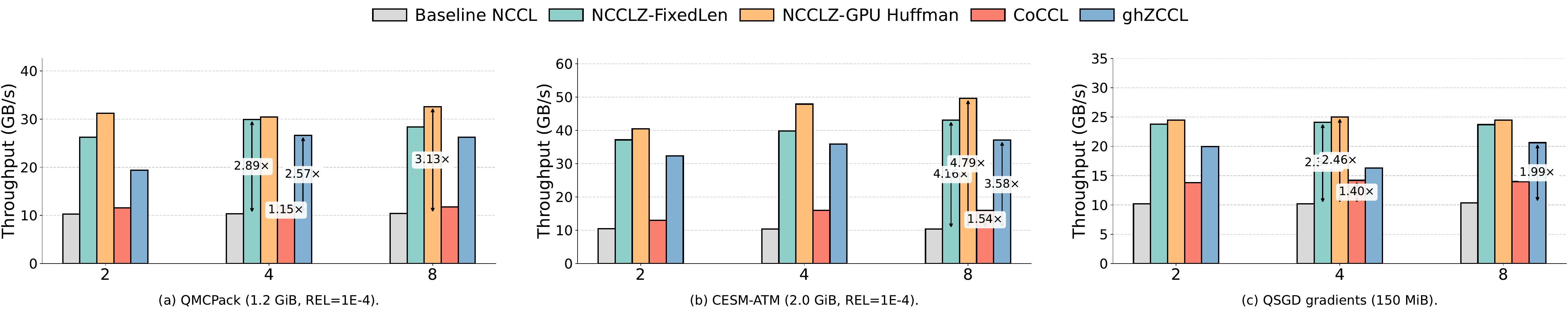}
    \caption{End-to-end \texttt{AllReduce} throughput on three real-world datasets using four GPUs per node. The x-axis shows nodes; the methods are unmodified NCCL, CoCCL, ghZCCL, NCCLZ-\textsc{FixedLen}, and NCCLZ-GPU \textsc{Huffman}.}
    \label{fig:e2e-realworld}
        \vspace{-9pt}
\end{figure*}

\subsection{End-to-End Performance on Real Datasets}
We evaluate end-to-end \texttt{AllReduce} throughput on three real-world tensors: two scientific datasets with REL=$10^{-4}$ and one training-gradient tensor quantized with QSGD. We compare NCCLZ with \textsc{FixedLen} or \textsc{GPU Huffman} against unmodified NCCL, CoCCL, and ghZCCL.

As shown in Figure~\ref{fig:e2e-realworld}, \sys achieves the highest end-to-end \texttt{AllReduce} throughput among the compared methods for all three datasets and node counts.
On QMCPack, \sys raises throughput from roughly 10\,GB/s to about 26--33\,GB/s, reaching up to 2.89$\times$ with \textsc{FixedLen} and 3.13$\times$ with \textsc{GPU Huffman}.
On QSGD gradients, \sys still improves throughput to about 24--25\,GB/s, achieving up to 2.37$\times$ with \textsc{FixedLen} and 2.46$\times$ with \textsc{GPU Huffman}.
In all cases, \sys outperforms CoCCL and ghZCCL. The advantage of \textsc{GPU Huffman} over \textsc{FixedLen} is most pronounced on CESM-ATM, whose symbol stream has lower entropy and greater compressibility.

Two effects contribute to these gains. First, quantization and entropy coding reduce the number of bytes entering NCCL's network path. Second, the native NCCL schedule allows encoding and decoding to overlap communication rather than run as a separate serialized stage. The relative contribution depends on the message and data distribution: byte reduction dominates when the transfer is bandwidth bound, while overlap determines how much codec cost remains exposed. \textsc{FixedLen} has low, predictable overhead; \textsc{GPU Huffman} is preferable only when its additional byte reduction amortizes its higher cost, as on CESM-ATM.

\subsection{Visual Quality Study}
We evaluate the reconstruction quality of \sys on QMCPack using
the same deterministic error-bounded quantizer as in the preceding
experiments, with REL=$10^{-4}$. Because unmodified NCCL transmits
the original FP32 payload without lossy approximation, its output
serves as the lossless reference. Figure~\ref{fig:qmcpack} compares
the 700th slice along the third dimension with the corresponding
output reconstructed by \sys. Both panels use identical color limits,
and the circular insets magnify the same spatial region.

At the plotted scale, the NCCLZ reconstruction preserves the
large-scale spatial distribution of the field, including the locations
and shapes of the warm and cool regions. The magnified view also
retains the local gradients and cell-level transition pattern, with no
visually discernible artifacts at REL=$10^{-4}$. This comparison
characterizes the end-to-end distortion of NCCLZ relative to the
lossless NCCL reference.

The observed distortion, if any, originates exclusively from the
one-time quantization performed at the interface. Both
\textsc{FixedLen} and GPU \textsc{Huffman} encode the resulting
symbol stream losslessly and therefore do not introduce additional
numerical error. Consequently, the two entropy coders produce the
same reconstruction quality at a fixed quantization setting; their
differences lie in compression ratio and execution cost. As reported
in Table~\ref{tab:cr_twocol_quant_fixed_huff_no_overall}, entropy
coding increases the average QMCPack compression ratio from
3.05$\times$ for quantization alone to 4.63$\times$ with
\textsc{FixedLen} and 8.25$\times$ with GPU \textsc{Huffman},
without changing the quantization-induced error.

\subsection{\sys~Scalability Study}

\begin{figure}[t]
  \centering
  \includegraphics[width=0.99\linewidth]{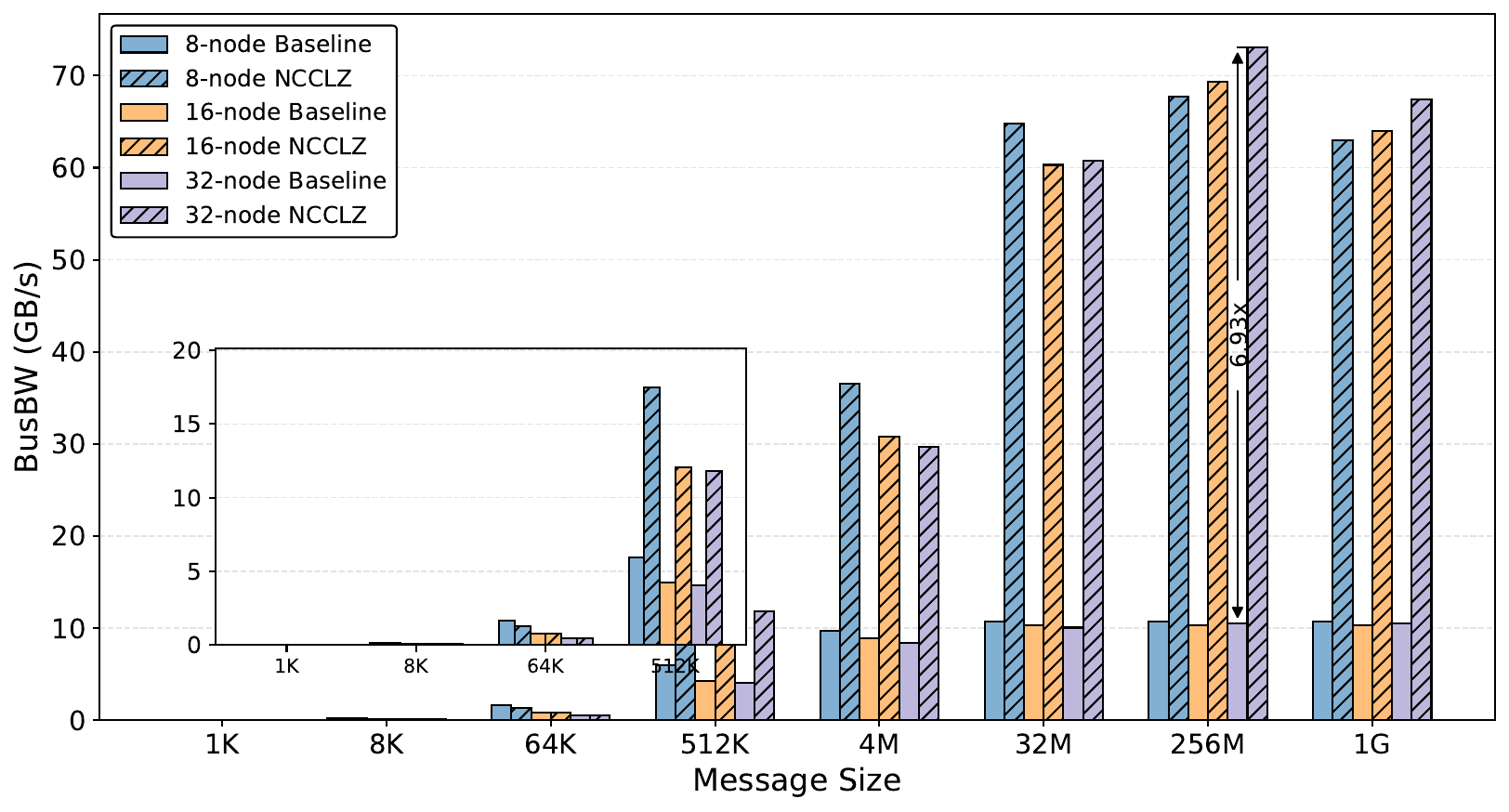}
  \caption{\texttt{AllToAll} BusBW of unmodified NCCL and NCCLZ versus message size on 8, 16, and 32 nodes.}
  \vspace{-6pt}
  \label{fig:scale-allgather}
\end{figure}

\begin{figure}[t]
  \centering
  \includegraphics[width=0.99\linewidth]{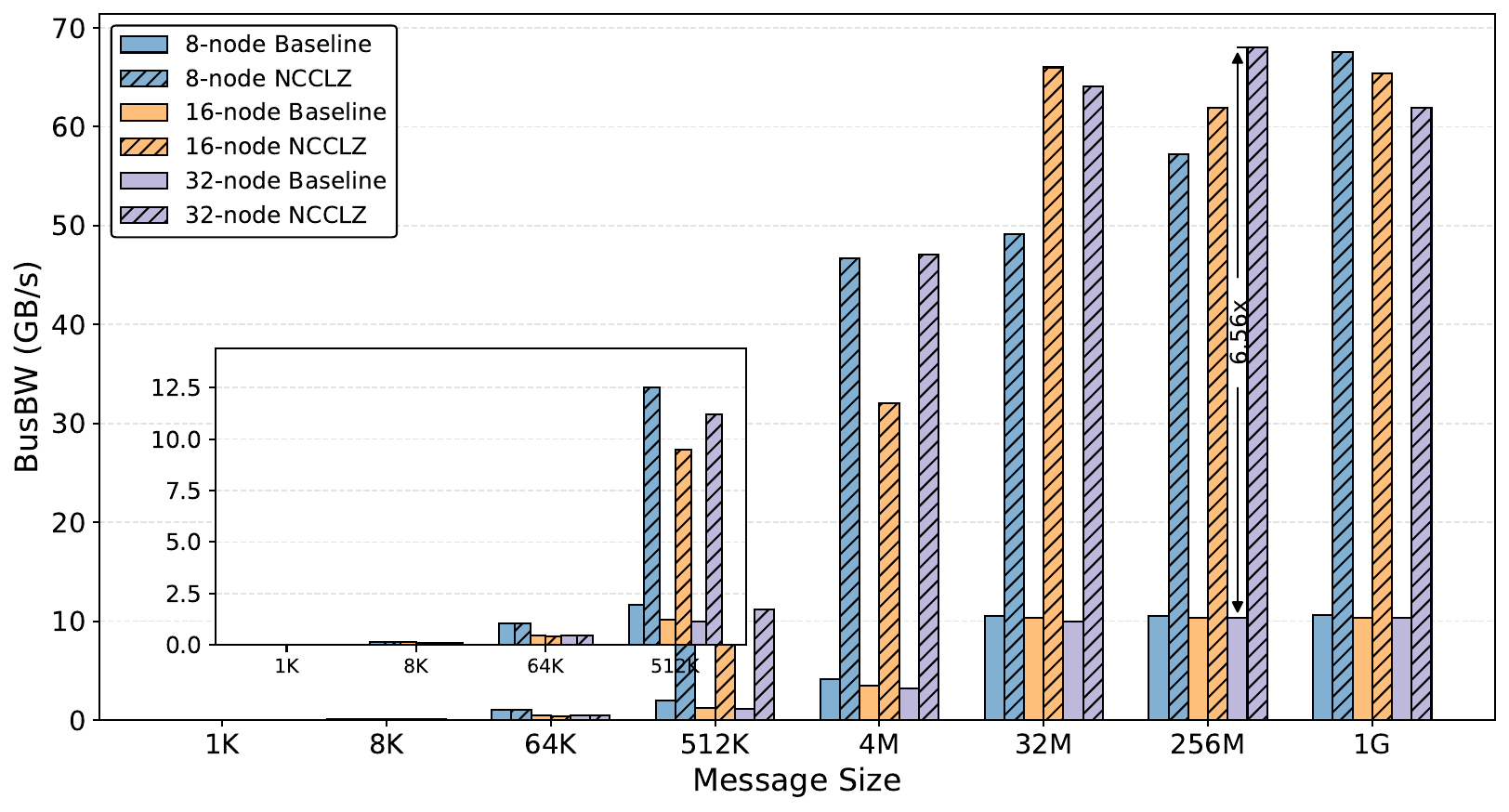}
  \caption{\texttt{AllReduce} BusBW of unmodified NCCL and NCCLZ versus message size on 8, 16, and 32 nodes.}
  \vspace{-8pt}
  \label{fig:scale-allreduce}
\end{figure}

We sweep message sizes for \texttt{AllToAll} and \texttt{AllReduce} on 8, 16, and 32 nodes. Figures~\ref{fig:scale-allgather} and~\ref{fig:scale-allreduce} compare NCCLZ with unmodified NCCL; they are intended to isolate scale-out behavior rather than provide a cross-system comparison. For messages up to 64\,KiB, fixed costs dominate and the difference is small. From 512\,KiB onward, reducing the transmitted bytes increases BusBW: unmodified NCCL remains near or below 10\,GB/s, whereas NCCLZ reaches tens of GB/s and peaks near 68\,GB/s, with a maximum improvement of 6.56$\times$. The trend persists from 8 to 32 nodes in both collectives, indicating that the gain is associated with network-payload reduction rather than a particular node count.

\subsection{\sys~Ablation Study}
\label{sec:eval-overlap}
We use ablation experiments to separate the benefit of additional byte reduction from the benefit of overlapping coding with communication.

\begin{figure}[!b]
    \centering
    \includegraphics[width=0.9\linewidth]{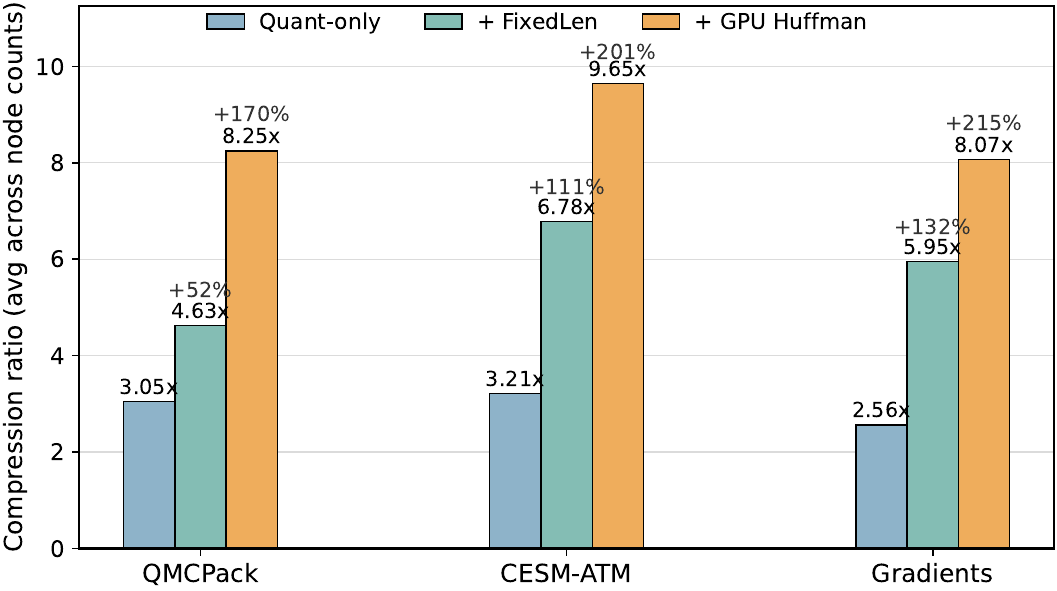}
    \caption{Average compression ratio across node counts for quantization alone and for quantization followed by \textsc{FixedLen} or GPU \textsc{Huffman}. Percentages above the entropy-coded bars give the improvement over quantization alone.}
    \label{fig:entropy-benefit}
    \vspace{-9pt}
\end{figure}

\textbf{Entropy Coding Benefit.}
Figure~\ref{fig:entropy-benefit} compares the average compression ratio across node counts for three workloads. The ratio is measured relative to the raw, pre-quantization size. Each workload uses quantization alone, quantization followed by \textsc{FixedLen}, or quantization followed by GPU \textsc{Huffman}.

Entropy coding provides a second reduction stage beyond quantization alone. \textsc{FixedLen} improves the ratio for all three workloads, and GPU \textsc{Huffman} provides the largest byte reduction, although the magnitude depends on the post-quantization symbol distribution. Because additional compression does not always offset codec cost, REA selects the path whose predicted reduction is most likely to improve end-to-end time.

\textbf{Overlap Efficiency.} We compare measured NCCLZ end-to-end time with a synthetic \emph{no-overlap} baseline in which coding and communication are serialized. For each message size, we measure encode and decode time separately and add both to the communication component to construct the no-overlap bar. The NCCLZ bar is the measured execution with its normal pipeline enabled; it includes all encode and decode overhead that remains exposed.

\begin{figure}[!t]
    \centering
    \includegraphics[width=0.98\columnwidth]{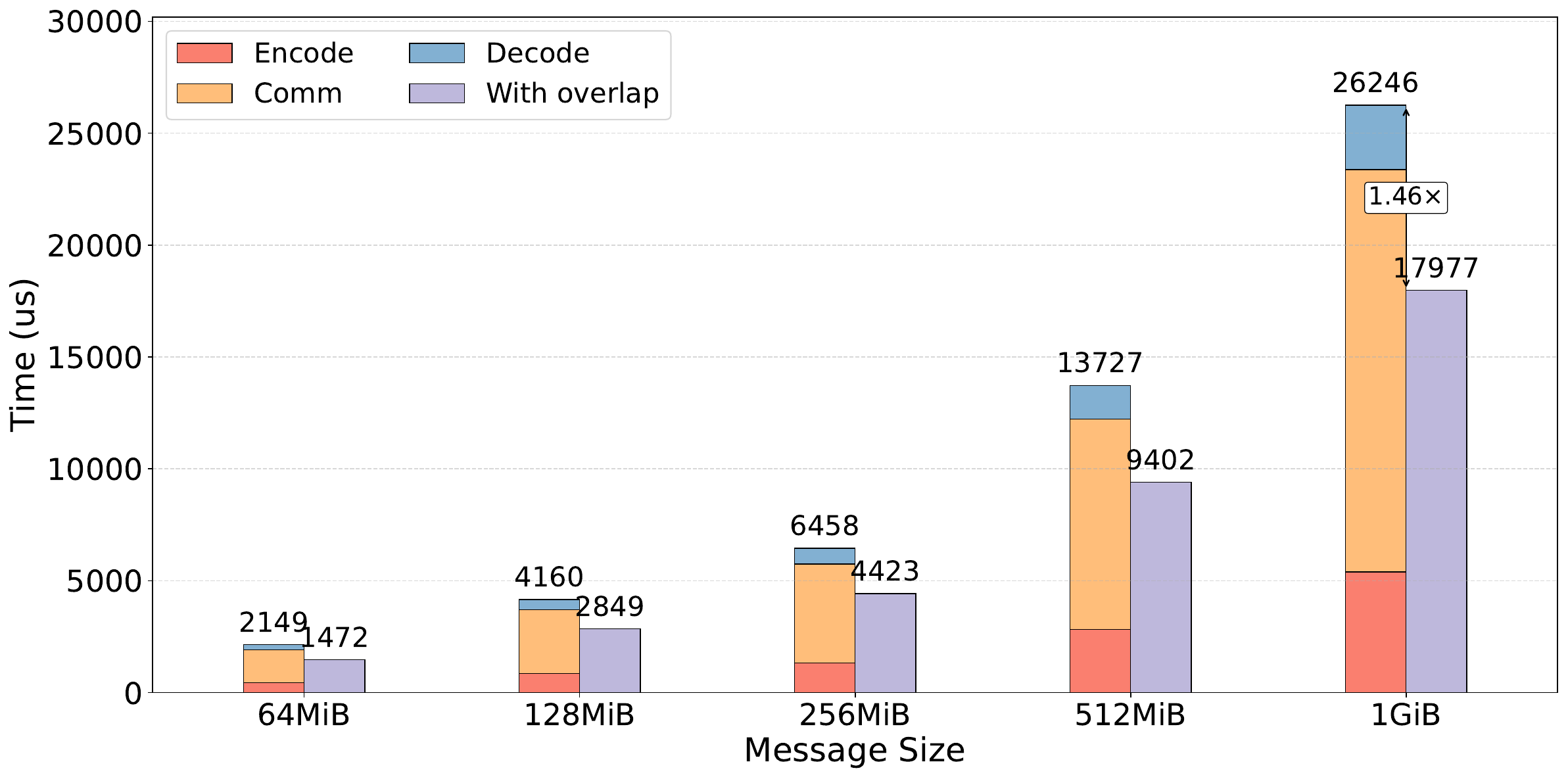}
    \caption{End-to-end time with NCCLZ overlap and with serialized encoding, communication, and decoding for 64\,MiB--1\,GiB messages. The 1.46$\times$ annotation marks only the maximum reduction, at 1\,GiB.}
    \vspace{-10pt}
    \label{fig:rs-overlap-vs-nooverlap}
\end{figure}

Figure~\ref{fig:rs-overlap-vs-nooverlap} shows that overlap reduces end-to-end time from 64\,MiB to 1\,GiB. The 1.46$\times$ annotation applies only to the 1-GiB result, where the reduction is largest; it does not indicate a constant speedup across message sizes. The gap between each pair of bars is the effective pipelining benefit at that size.

As message size grows, the encode component of the serialized baseline increases because Huffman bitstream construction and global-memory traffic also increase. Less communication slack is then available relative to the codec work, so a larger fraction of encoding can remain exposed. Nevertheless, NCCLZ remains faster than the serialized baseline at 512\,MiB and 1\,GiB, showing that slot-level pipelining hides a substantial portion of the coding cost in the bandwidth-bound regime.

\section{Related Work} \label{sec:related}
\textbf{GPU Compression and Quantization.}
GPU-resident data reduction mitigates bandwidth bottlenecks without CPU staging. Scientific compressors such as SZ combine prediction, error-controlled quantization, and entropy coding~\cite{di2016sz}; ZFP uses a block transform with fixed-rate and fixed-accuracy modes~\cite{lindstrom2014fixedrate}. cuSZ adapts SZ-style compression and entropy coding to GPUs~\cite{tian2020cusz}, and ZFP supports CUDA-based whole-array compression and decompression~\cite{llnl_zfp}. Distributed deep learning more commonly uses communication-oriented gradient quantization, often with error feedback, rather than strict pointwise error bounds~\cite{seide2014onebit}.

\textbf{Compression in MPI-Based Collectives.}
MPI compression predates GPU-resident CCLs. AdOC and cMPI introduced transparent runtime compression~\cite{Jeannot2005AdOC,Ke2004cMPI}, while CoMPI and PRAcTICaL-MPI select codecs at runtime from message and network characteristics~\cite{Filgueira2009CoMPI,Filgueira2012PRAcTICaL}. The Panda line co-designs online GPU compression with MPI collective algorithms to improve overlap~\cite{Zhou2022AlltoAllOnlineCompression,Zhou2024ISCAllReduceCompression}, and C-Coll integrates error-bounded lossy compression into MPI collectives~\cite{Huang2023CColl}. These systems demonstrate that runtime adaptation, GPU acceleration, and controlled error are feasible in MPI. NCCLZ does not assume that stage separation is impossible in MPI; its distinction is to separate quantization from entropy coding at NCCL-specific interface and primitive boundaries so the lossless stage can use NCCL's device pipeline.

\textbf{Collective Communication Libraries.}
MPI is a general-purpose collective substrate with portable, highly tuned implementations such as MPICH and MVAPICH2~\cite{mpich_overview,gropp1996mpich,huang2006mvapich2}. GPU-oriented variants such as MVAPICH2-GDR add GPUDirect RDMA and optimized device-buffer collectives~\cite{mvapich2gdr_userguide}. Vendor and framework CCLs expose more GPU-specific interfaces: RCCL targets AMD GPUs~\cite{amd_rccl_doc}, oneCCL targets Intel and heterogeneous environments~\cite{oneccl_docs}, and Gloo provides a multi-backend collective layer used by PyTorch~\cite{gloo_repo}. NCCLZ is implemented for NCCL, but its separation of application-controlled quantization from in-runtime lossless coding can also inform other CCLs; applying the design to those libraries remains future work.

\section{Conclusion and Future Work}
We presented NCCLZ, a compression-enabled NCCL framework for bandwidth-constrained GPU clusters. NCCLZ places lossy quantization at the interface and lossless entropy coding inside NCCL primitives, allowing the two stages to be selected independently while the in-runtime stage overlaps communication. Across scientific data, training gradients, and synthetic workloads, NCCLZ improves compression ratio and end-to-end collective throughput, with gains of up to $9.65\times$ over unmodified NCCL and $3.34\times$ over existing compression-assisted CCLs. Future work will extend the design to other CCLs and tune it for sharded data parallelism and collective-intensive LLM parallelism.

\section*{Acknowledgment}
This work was supported by the U.S. National Science Foundation under award OAC-2348465. This research used resources of the Argonne Leadership Computing Facility, which is a U.S. Department of Energy Office of Science User Facility operated under contract DE-AC02-06CH11357.

AI was used only for language polishing and for
checking grammar, spelling, and formatting in this manuscript.
All technical content, results, claims, and references were
reviewed and verified by the authors.

\bibliographystyle{IEEEtran}
\bibliography{references}

@article{Hu2025DemystifyingNCCL,
  title   = {Demystifying NCCL: An In-depth Analysis of {GPU} Communication Protocols and Algorithms},
  author  = {Hu, Zhiyi and Shen, Siyuan and Bonato, Tommaso and Jeaugey, Sylvain and Alexander, Cedell and Spada, Eric and Dinan, James and Hammond, Jeff and Hoefler, Torsten},
  journal = {arXiv preprint arXiv:2507.04786},
  year    = {2025},
  doi     = {10.48550/arXiv.2507.04786},
  url     = {https://doi.org/10.48550/arXiv.2507.04786},
  volume  = {abs/2507.04786},
  pages   = {1--24},
}

@inproceedings{Huang2025ghZCCL,
  title     = {ghZCCL: Advancing {GPU}-aware Collective Communications with Homomorphic Compression},
  author    = {Huang, Jiajun and Di, Sheng and Huang, Yafan and Chen, Zizhong and Cappello, Franck and Guo, Yanfei and Thakur, Rajeev},
  booktitle = {Proceedings of the 2025 International Conference on Supercomputing},
  series    = {ICS '25},
  year      = {2025},
  address   = {New York, NY, USA},
  publisher = {Association for Computing Machinery},
  numpages  = {14},
  doi       = {10.1145/3721145.3733642},
  url       = {https://doi.org/10.1145/3721145.3733642},
}

@misc{NCCL,
  title        = {{NVIDIA} Collective Communications Library ({NCCL})},
  author       = {{NVIDIA}},
  url         = {https://github.com/NVIDIA/nccl},
  year         = {2025},
  note         = {Accessed: 2026-01-16},
}

@inproceedings{Zhang2020NetBottleneck,
  title     = {Is Network the Bottleneck of Distributed Training?},
  author    = {Zhang, Zhen and Chang, Chaokun and Lin, Haibin and Wang, Yida and Arora, Raman and Jin, Xin},
  booktitle = {Workshop on Network Meets {AI} \& {ML}},
  series    = {NetAI '20},
  year      = {2020},
  address   = {Virtual Event, NY, USA},
  publisher = {Association for Computing Machinery},
  numpages  = {6},
  doi       = {10.1145/3405671.3405810},
  url       = {https://doi.org/10.1145/3405671.3405810},
}

@misc{COCCLRepo,
  title        = {{COCCL}: Compression and Precision Co-aware Collective Communication Library},
  author       = {Liu, Xingchen and Kong, Haoran and Wei, Zheng and Zhao, Liyang and Wang, Yufan and Yang, Jinwu},
  url         = {https://github.com/hpdps-group/COCCL},
  year         = {2025},
  note         = {Accessed: 2026-01-26},
}

@misc{Liang2024CommEffSurvey,
  title         = {Communication-Efficient Large-Scale Distributed Deep Learning: A Comprehensive Survey},
  author        = {Liang, Feng and Zhang, Zhen and Lu, Haifeng and Leung, Victor C. M. and Guo, Yanyi and Hu, Xiping},
  year          = {2024},
  eprint        = {2404.06114},
  archivePrefix = {arXiv},
  primaryClass  = {cs.DC},
  doi           = {10.48550/arXiv.2404.06114},
  url           = {https://doi.org/10.48550/arXiv.2404.06114},
}

@inproceedings{Poseidon2017ATC,
  title     = {Poseidon: An Efficient Communication Architecture for Distributed Deep Learning on GPU Clusters},
  author    = {Zhang, Hao and Zheng, Zeyu and Xu, Shizhen and Dai, Wei and Ho, Qirong and Liang, Xiaodan and Hu, Zhiting and Wei, Jinliang and Xie, Pengtao and Xing, Eric P.},
  booktitle = {2017 USENIX Annual Technical Conference (USENIX ATC 17)},
  pages     = {181--193},
  year      = {2017},
  publisher = {USENIX Association},
  address   = {Santa Clara, CA, USA},
  url       = {https://www.usenix.org/conference/atc17/technical-sessions/presentation/zhang},
}

@article{Sergeev2018Horovod,
  title   = {Horovod: fast and easy distributed deep learning in TensorFlow},
  author  = {Sergeev, Alexander and Del Balso, Mike},
  journal = {arXiv preprint arXiv:1802.05799},
  year    = {2018},
  doi     = {10.48550/arXiv.1802.05799},
  url     = {https://doi.org/10.48550/arXiv.1802.05799},
  volume  = {abs/1802.05799},
  pages   = {1--13},
}

@article{Huang2023CColl,
  title   = {C-Coll: Introducing Error-bounded Lossy Compression into MPI Collectives},
  author  = {Huang, Jiajun and Di, Sheng and Yu, Xiaodong and Zhai, Yujia and Liu, Jinyang and Raffenetti, Ken and Zhou, Hui and Zhao, Kai and Chen, Zizhong and Cappello, Franck and Guo, Yanfei and Thakur, Rajeev},
  journal = {arXiv preprint arXiv:2304.03890},
  year    = {2023},
  doi     = {10.48550/arXiv.2304.03890},
  url     = {https://doi.org/10.48550/arXiv.2304.03890},
  volume  = {abs/2304.03890},
  pages   = {1--19},
}

@inproceedings{Zhou2021MVAPICHCompression,
  title     = {Designing High-Performance {MPI} Libraries with On-the-fly Compression for Modern {GPU} Clusters},
  author    = {Zhou, Qinghua and Chu, Ching-Hsiang and Kumar, N. Senthil and Kousha, Pouya and Ghazimirsaeed, Seyedeh Mahdieh and Subramoni, Hari and Panda, Dhabaleswar K.},
  booktitle = {2021 IEEE International Parallel and Distributed Processing Symposium (IPDPS)},
  year      = {2021},
  pages     = {444--453},
  doi       = {10.1109/IPDPS49936.2021.00053},
  url       = {https://doi.org/10.1109/IPDPS49936.2021.00053},
  publisher = {IEEE},
  address   = {Portland, OR, USA},
}

@inproceedings{Zhou2022AlltoAllOnlineCompression,
  title     = {Accelerating {MPI} All-to-All Communication with Online Compression on Modern {GPU} Clusters},
  author    = {Zhou, Qinghua and Kousha, Pouya and Anthony, Quentin and Khorassani, Kawthar Shafie and Shafi, Aamir and Subramoni, Hari and Panda, Dhabaleswar K.},
  booktitle = {High Performance Computing -- 37th International Conference, ISC High Performance 2022, Hamburg, Germany, May 29--June 2, 2022, Proceedings},
  editor    = {Varbanescu, Ana Lucia and Bhatele, Abhinav and Luszczek, Piotr and Baboulin, Marc},
  series    = {Lecture Notes in Computer Science},
  volume    = {13289},
  pages     = {3--25},
  publisher = {Springer},
  year      = {2022},
  doi       = {10.1007/978-3-031-07312-0_1},
  url       = {https://doi.org/10.1007/978-3-031-07312-0_1},
  address   = {Hamburg, Germany},
}

@inproceedings{Zhou2024ISCAllReduceCompression,
  author    = {Qinghua Zhou and,
                  Bharath Ramesh and
                  Aamir Shafi and
                  Mustafa Abduljabbar and
                  Hari Subramoni and
                  Dhabaleswar K. Panda},
  title     = {Accelerating {MPI} AllReduce Communication with Efficient GPU-Based,
                  Compression Schemes on Modern {GPU} Clusters},
  booktitle = {{ISC} High Performance 2024 Research Paper Proceedings (39th International,
                  Conference), Hamburg, Germany, May 12-16, 2024},
  pages     = {1--12},
  publisher = {Prometeus GmbH / {IEEE}},
  year      = {2024},
  doi       = {10.23919/ISC.2024.10528931},
  url       = {https://doi.org/10.23919/ISC.2024.10528931},
  address   = {Hamburg, Germany},
}

@inproceedings{Huang2024gZCCL,
  title     = {g{ZCCL}: Compression-Accelerated Collective Communication Framework for {GPU} Clusters},
  author    = {Huang, Jiajun and Di, Sheng and Yu, Xiaodong and Zhai, Yujia and Liu, Jinyang and Huang, Yafan and Raffenetti, Ken and Zhou, Hui and Zhao, Kai and Lu, Xiaoyi and Chen, Zizhong and Cappello, Franck and Guo, Yanfei and Thakur, Rajeev},
  booktitle = {Proceedings of the 38th ACM International Conference on Supercomputing (ICS '24)},
  year      = {2024},
  pages     = {437--448},
  doi       = {10.1145/3650200.3656636},
  url       = {https://doi.org/10.1145/3650200.3656636},
  publisher = {Association for Computing Machinery},
  address   = {New York, NY, USA},
}

@article{Di2018PointwiseSZ,
  title   = {Efficient Lossy Compression for Scientific Data Based on Pointwise Relative Error Bound},
  author  = {Di, Sheng and Tao, Dingwen and Liang, Xin and Cappello, Franck},
  journal = {IEEE Transactions on Parallel and Distributed Systems},
  volume  = {30},
  number  = {2},
  pages   = {331--345},
  year    = {2018},
  doi     = {10.1109/TPDS.2018.2859932},
  url     = {https://doi.org/10.1109/TPDS.2018.2859932},
}

@article{Lindstrom2014ZFP,
  title   = {Fixed-Rate Compressed Floating-Point Arrays},
  author  = {Lindstrom, Peter},
  journal = {IEEE Transactions on Visualization and Computer Graphics},
  volume  = {20},
  number  = {12},
  pages   = {2674--2683},
  year    = {2014},
  doi     = {10.1109/TVCG.2014.2346458},
  url     = {https://doi.org/10.1109/TVCG.2014.2346458},
}

@inproceedings{Huang2023CuSZp,
  title     = {{CuSZp}: An Ultra-fast {GPU} Error-bounded Lossy Compression Framework with Optimized End-to-End Performance},
  author    = {Huang, Yafan and Di, Sheng and Yu, Xiaodong and Li, Guanpeng and Cappello, Franck},
  booktitle = {Proceedings of the International Conference for High Performance Computing, Networking, Storage and Analysis},
  series    = {SC '23},
  year      = {2023},
  publisher = {Association for Computing Machinery},
  address   = {New York, NY, USA},
  doi       = {10.1145/3581784.3607048},
  url       = {https://doi.org/10.1145/3581784.3607048},
  articleno = {43},
  numpages  = {13},
}

@article{Alistarh2017QSGD,
  title   = {{QSGD}: Communication-Efficient {SGD} via Gradient Quantization and Encoding},
  author  = {Alistarh, Dan and Grubic, Demjan and Li, Jerry and Tomioka, Ryota and Vojnovic, Milan},
  journal = {arXiv preprint arXiv:1610.02132},
  year    = {2017},
  doi     = {10.48550/arXiv.1610.02132},
  url     = {https://doi.org/10.48550/arXiv.1610.02132},
  volume  = {abs/1610.02132},
  pages   = {1--14},
}

@misc{NvidiaNCCLDev,
  title        = {{NVIDIA} Collective Communications Library ({NCCL})},
  author       = {{NVIDIA}},
  url         = {https://developer.nvidia.com/nccl},
  year         = {2026},
  note         = {Accessed: 2026-01-27},
}

@misc{PyTorchDDP,
  title        = {DistributedDataParallel --- {PyTorch} Documentation},
  author       = {{PyTorch}},
  url         = {https://docs.pytorch.org/docs/stable/generated/torch.nn.parallel.DistributedDataParallel.html},
  year         = {2026},
  note         = {Accessed: 2026-01-27},
}

@article{Shoeybi2019MegatronLM,
  title   = {Megatron-LM: Training Multi-Billion Parameter Language Models Using Model Parallelism},
  author  = {Shoeybi, Mohammad and Patwary, Mostofa and Puri, Raul and LeGresley, Patrick and Casper, Jared and Catanzaro, Bryan},
  journal = {arXiv preprint arXiv:1909.08053},
  year    = {2019},
  doi     = {10.48550/arXiv.1909.08053},
  url     = {https://doi.org/10.48550/arXiv.1909.08053},
  volume  = {abs/1909.08053},
  pages   = {1--12},
}

@article{Kwon2023PagedAttention,
  title   = {Efficient Memory Management for Large Language Model Serving with {PagedAttention}},
  author  = {Kwon, Woosuk and Li, Zhuohan and Zhuang, Siyuan and Sheng, Ying and Zheng, Lianmin and Yu, Cody Hao and Gonzalez, Joseph E. and Zhang, Hao and Stoica, Ion},
  journal = {arXiv preprint arXiv:2309.06180},
  year    = {2023},
  doi     = {10.48550/arXiv.2309.06180},
  url     = {https://doi.org/10.48550/arXiv.2309.06180},
  volume  = {abs/2309.06180},
  pages   = {1--17},
}

@article{Zheng2024SGLang,
  title   = {{SGLang}: Efficient Execution of Structured Language Model Programs},
  author  = {Zheng, Lianmin and Yin, Liangsheng and Xie, Zhiqiang and Sun, Chuyue and Huang, Jeff and Yu, Cody Hao and Cao, Shiyi and Kozyrakis, Christos and Stoica, Ion and Gonzalez, Joseph E. and Barrett, Clark and Sheng, Ying},
  journal = {arXiv preprint arXiv:2312.07104},
  year    = {2024},
  doi     = {10.48550/arXiv.2312.07104},
  url     = {https://doi.org/10.48550/arXiv.2312.07104},
  volume  = {abs/2312.07104},
  pages   = {1--16},
}

@techreport{NVIDIAA100NVLink,
  title       = {{NVIDIA A100 80GB PCIe GPU} (Product Brief)},
  author      = {{NVIDIA}},
  institution = {{NVIDIA}},
  year        = {2022},
  note        = {Reports up to 600~GB/s NVLink bandwidth with NVLink bridges.},
  url         = {https://www.nvidia.com/content/dam/en-zz/Solutions/Data-Center/a100/pdf/PB-10577-001_v02.pdf},
}

@misc{alcf-polaris,
  title        = {Polaris},
  howpublished = {Argonne Leadership Computing Facility (ALCF)},
  url          = {https://www.alcf.anl.gov/polaris},
  note         = {Accessed: 2026-02-03},
  author       = {Argonne Leadership Computing Facility},
  year         = {2026},
}

@inproceedings{ZhouCKKGS021,
  author    = {Qinghua Zhou and C. Chu and N. S. Kumar and Pouya Kousha and,
               Seyedeh Mahdieh Ghazimirsaeed and Hari Subramoni and Dhabaleswar K. Panda},
  title     = {Designing High-Performance {MPI} Libraries with On-the-fly Compression for Modern {GPU} Clusters},
  booktitle = {35th {IEEE} International Parallel and Distributed Processing Symposium, {IPDPS} 2021,
               Portland, OR, USA, May 17--21, 2021},
  pages     = {444--453},
  publisher = {{IEEE}},
  year      = {2021},
  doi       = {10.1109/IPDPS49936.2021.00053},
  url       = {https://doi.org/10.1109/IPDPS49936.2021.00053},
  address   = {Portland, OR, USA},
}

@inproceedings{Jeannot2005AdOC,
  author    = {Emmanuel Jeannot and Peter Strazdins},
  title     = {Improving Middleware Performance with {AdOC}: An Adaptive Online Compression Library for Data Transfer},
  booktitle = {Proceedings of the 19th {IEEE} International Parallel and Distributed Processing Symposium ({IPDPS})},
  year      = {2005},
  publisher = {IEEE},
  doi       = {10.1109/IPDPS.2005.254},
  url       = {https://doi.org/10.1109/IPDPS.2005.254},
  pages     = {1--8},
  address   = {Denver, CO, USA},
}

@inproceedings{Ke2004cMPI,
  author    = {Jian Ke and Martin Burtscher and Evan Speight},
  title     = {Runtime Compression of {MPI} Messages to Improve the Performance and Scalability of Parallel Applications},
  booktitle = {Proceedings of the {ACM/IEEE} Conference on Supercomputing (SC '04)},
  year      = {2004},
  pages     = {59},
  publisher = {IEEE Computer Society},
  address   = {Pittsburgh, PA, USA},
  doi       = {10.1109/SC.2004.52},
  url       = {https://doi.org/10.1109/SC.2004.52},
}

@inproceedings{Filgueira2009CoMPI,
  author    = {Rosa Filgueira and David E. Singh and Alejandro Calder{\'o}n and Jes{\'u}s Carretero},
  title     = {CoMPI: Enhancing {MPI} Based Applications Performance and Scalability Using Run-Time Compression},
  booktitle = {Recent Advances in Parallel Virtual Machine and Message Passing Interface (EuroPVM/MPI 2009)},
  series    = {Lecture Notes in Computer Science},
  volume    = {5759},
  pages     = {207--218},
  year      = {2009},
  publisher = {Springer},
  doi       = {10.1007/978-3-642-03770-2_27},
  url       = {https://doi.org/10.1007/978-3-642-03770-2_27},
  address   = {Espoo, Finland},
}

@inproceedings{Filgueira2012PRAcTICaL,
  author    = {Rosa Filgueira and Malcolm Atkinson and Alberto Nu{\~n}ez and Javier Fern{\'a}ndez},
  title     = {An Adaptive, Scalable, and Portable Technique for Speeding Up {MPI}-Based Applications},
  booktitle = {Euro-Par 2012 Parallel Processing},
  series    = {Lecture Notes in Computer Science},
  volume    = {7484},
  pages     = {729--740},
  year      = {2012},
  publisher = {Springer},
  doi       = {10.1007/978-3-642-32820-6_72},
  url       = {https://doi.org/10.1007/978-3-642-32820-6_72},
  address   = {Rhodes Island, Greece},
}

@inproceedings{seide2014onebit,
  title     = {1-Bit Stochastic Gradient Descent and its Application to Data-Parallel Distributed Training of Speech {DNN}s},
  author    = {Seide, Frank and Fu, Hao and Droppo, Jasha and Li, Gang and Yu, Dong},
  booktitle = {INTERSPEECH},
  year      = {2014},
  url       = {https://www.microsoft.com/en-us/research/wp-content/uploads/2016/02/IS140694.pdf},
  pages     = {1058--1062},
  publisher = {ISCA},
  address   = {Singapore},
}

@inproceedings{di2016sz,
  title     = {Fast Error-bounded Lossy {HPC} Data Compression with {SZ}},
  author    = {Di, Sheng and Cappello, Franck},
  booktitle = {2016 IEEE International Parallel and Distributed Processing Symposium (IPDPS)},
  pages     = {730--739},
  year      = {2016},
  url       = {https://szcompressor.org/tabs/publication/},
  publisher = {IEEE},
  address   = {Chicago, IL, USA},
}

@article{lindstrom2014fixedrate,
  title   = {Fixed-Rate Compressed Floating-Point Arrays},
  author  = {Lindstrom, Peter},
  journal = {IEEE Transactions on Visualization and Computer Graphics},
  volume  = {20},
  number  = {12},
  pages   = {2674--2683},
  year    = {2014},
  doi     = {10.1109/TVCG.2014.2346458},
  url     = {https://doi.org/10.1109/TVCG.2014.2346458},
}

@inproceedings{tian2020cusz,
  title     = {cuSZ: An Efficient {GPU}-Based Error-Bounded Lossy Compression Framework for Scientific Data},
  author    = {Tian, Jiannan and Di, Sheng and Zhao, Kai and Rivera, Cody and Hickman Fulp, Megan and Underwood, Robert and Jin, Sian and Liang, Xin and Calhoun, Jon and Tao, Dingwen and Cappello, Franck},
  booktitle = {Proceedings of the 29th International Conference on Parallel Architectures and Compilation Techniques (PACT)},
  year      = {2020},
  doi       = {10.1145/3410463.3414624},
  url       = {https://doi.org/10.1145/3410463.3414624},
  pages     = {1--12},
  publisher = {Association for Computing Machinery},
  address   = {New York, NY, USA},
}

@article{gropp1996mpich,
  author  = {Gropp, William and Lusk, Ewing and Doss, Nathan and Skjellum, Anthony},
  title   = {A High-Performance, Portable Implementation of the MPI Message Passing Interface Standard},
  journal = {Parallel Computing},
  year    = {1996},
  doi     = {10.1016/0167-8191(96)00024-5},
  url     = {https://doi.org/10.1016/0167-8191(96)00024-5},
  volume  = {22},
  number  = {6},
  pages   = {789--828},
}

@inproceedings{huang2006mvapich2,
  author    = {Huang, Wei and Santhanaraman, Gopalakrishnan and Jin, Hyun-Wook and Gao, Qi and Panda, Dhabaleswar K.},
  title     = {Design of High Performance MVAPICH2: MPI2 over InfiniBand},
  booktitle = {Proceedings of the Sixth IEEE International Symposium on Cluster Computing and the Grid (CCGRID)},
  year      = {2006},
  doi       = {10.1109/CCGRID.2006.32},
  url       = {https://doi.org/10.1109/CCGRID.2006.32},
  pages     = {43--48},
  publisher = {IEEE},
  address   = {Singapore},
}

@misc{mpich_overview,
  title        = {MPICH Overview},
  author       = {{MPICH Project}},
  url         = {https://www.mpich.org/about/overview/},
  note         = {Accessed: 2026-02-06},
  year         = {2026},
}

@misc{mvapich2gdr_userguide,
  title        = {MVAPICH2-GDR User Guide},
  author       = {{MVAPICH Project}},
  url         = {https://mvapich.cse.ohio-state.edu/userguide/gdr/},
  note         = {Accessed: 2026-02-06},
  year         = {2026},
}

@misc{amd_rccl_doc,
  title        = {RCCL Documentation (ROCm Communication Collectives Library)},
  author       = {{AMD}},
  url         = {https://rocmdocs.amd.com/projects/rccl/en/latest/index.html},
  note         = {Accessed: 2026-02-06},
  year         = {2026},
}

@misc{oneccl_docs,
  title        = {oneAPI Collective Communications Library (oneCCL) Documentation},
  author       = {{UXL Foundation}},
  url         = {https://uxlfoundation.github.io/oneCCL/index.html},
  note         = {Accessed: 2026-02-06},
  year         = {2026},
}

@misc{gloo_repo,
  title        = {Gloo: Collective Communications Library},
  author       = {{PyTorch}},
  url         = {https://github.com/pytorch/gloo},
  note         = {Accessed: 2026-02-06},
  year         = {2026},
}

@misc{llnl_zfp,
  author       = {{LLNL}},
  title        = {{zfp}: Compressed Floating-Point and Integer Arrays (CUDA support)},
  howpublished = {GitHub repository},
  note         = {Accessed 2026-02-06},
  url          = {https://github.com/LLNL/zfp},
  year         = {2026},
}

@article{Chen2025GreedyLore,
  author  = {Chuyan Chen and Yutong He and Pengrui Li and Weichen Jia and Kun Yuan},
  title   = {Greedy Low-Rank Gradient Compression for Distributed Learning with Convergence Guarantees},
  journal = {arXiv preprint arXiv:2507.08784},
  year    = {2025},
  url     = {https://doi.org/10.48550/arXiv.2507.08784}
}

@inproceedings{Feng2024DLRMCompression,
  author    = {Hao Feng and Boyuan Zhang and Fanjiang Ye and Min Si and Ching-Hsiang Chu and Jiannan Tian and Chunxing Yin and Summer Deng and Yuchen Hao and Pavan Balaji and Tong Geng and Dingwen Tao},
  title     = {Accelerating Communication in Deep Learning Recommendation Model Training with Dual-Level Adaptive Lossy Compression},
  booktitle = {SC24: International Conference for High Performance Computing, Networking, Storage and Analysis},
  year      = {2024},
  pages     = {1--16},
  url       = {https://doi.org/10.1109/SC41406.2024.00095}
}

@article{He2025TAHQuant,
  author  = {Guangxin He and Yuan Cao and Yutong He and Tianyi Bai and Kun Yuan and Binhang Yuan},
  title   = {TAH-QUANT: Effective Activation Quantization in Pipeline Parallelism over Slow Network},
  journal = {arXiv preprint arXiv:2506.01352},
  year    = {2025},
  url     = {https://doi.org/10.48550/arXiv.2506.01352}
}

@inproceedings{Liu2026COCCL,
  author    = {Xingchen Liu and Haoran Kong and Hairui Zhao and Shengkai Lyu and Zheng Wei and Man Liu and Xingjian Tian and Liyang Zhao and Zhuohan Chen and Fakang Wang and Zizhong Chen and Zhan Wang and Guangming Tan and Dingwen Tao},
  title     = {COCCL: A Collective Communication Library Supporting Easy Integration and Configuration of Customized Compression for Scalable LLM Training},
  booktitle = {Proceedings of the 31st ACM SIGPLAN Annual Symposium on Principles and Practice of Parallel Programming},
  series    = {PPoPP '26},
  year      = {2026},
  pages     = {384--397},
  url       = {https://doi.org/10.1145/3774934.3786432}
}

\end{document}